\documentclass[aps,prx,twocolumn,english,showpacs,showkeys,superscriptaddress,amssymb,amsfonts]{revtex4}
\usepackage[T1]{fontenc}
\usepackage[latin9]{inputenc}
\usepackage{amsmath}
\usepackage{dsfont}
\usepackage{amstext}
\usepackage{amsthm}

\usepackage{tocvsec2}

\usepackage{amssymb}
\usepackage{amsbsy}
\usepackage{amsthm}
\usepackage{epsfig}
\usepackage{framed}
\usepackage{graphicx}
\usepackage{bbm}
\usepackage{hyperref}
\usepackage{color}
\usepackage{multirow}

\newcommand{\Ref}[1]{Ref.~\onlinecite{#1}}

\newcommand{\bst}{{\boldsymbol{T}}}

\newcommand{\ie}{{\emph{i.e.~}}}
\makeatletter

\newcommand{\Rmnum}[1]{\expandafter\@slowromancap\romannumeral #1@}
\makeatother
\newcommand{\imth}{\hspace{1pt}\mathrm{i}\hspace{1pt}}

\newcommand{\eg}{{\emph{e.g.~}}}

\newcommand{\mbz}{{\mathbb{Z}}}
\newcommand{\bea}{\begin{eqnarray}}
\newcommand{\eea}{\end{eqnarray}}
\newcommand{\bpm}{\begin{pmatrix}}
\newcommand{\epm}{\end{pmatrix}}
\newcommand{\bal}{\begin{aligned}}
\newcommand{\eal}{\end{aligned}}

\newcommand{\calP}{\mathcal{P}}
\newcommand{\calU}{\mathcal{U}}
\newcommand{\calF}{\mathcal{F}}
\newcommand{\calH}{\mathcal{H}}
\newcommand{\calT}{\mathcal{T}}

\newcommand{\frakF}{\mathfrak{F}}
\newcommand{\frakU}{\mathfrak{U}}
\newcommand{\frakp}{\mathfrak{p}}
\newcommand{\frakq}{\mathfrak{q}}

\newtheorem{theorem}{Theorem}

\makeatother

\usepackage{babel}
\begin{document}
\title{Filling-enforced constraint on the quantized Hall conductivity on a periodic lattice}

\author{Yuan-Ming Lu}
\affiliation{Department of Physics, The Ohio State University, Columbus, Ohio 43210, USA}
\author{Ying Ran}
\affiliation{Department of Physics, Boston College, Chestnut Hill, MA 02467, USA}
\author{Masaki Oshikawa}
\affiliation{Institute for Solid State Physics,the University of Tokyo,
5-1-5 Kashiwanoha, Kashiwa-shi, Chiba 277-8581, Japan}
\affiliation{Kavli Institute for Theoretical Physics, University of California, Santa Barbara, CA 93106, USA}
\date{\today}

\begin{abstract}
We discuss quantum Hall effects in a gapped insulator on a periodic two-dimensional lattice. We derive a universal relation among the the quantized Hall conductivity, and charge and flux densities per physical unit cell. This follows from the magnetic translation symmetry and the large gauge invariance, and holds for a very general class of interacting many-body systems. It can be understood as a combination of Laughlin's gauge invariance argument and Lieb-Schultz-Mattis-type theorem. A variety of complementary arguments, based on a cut-and-glue procedure, the many-body electric polarization, and a fractionalization algebra of magnetic translation symmetry, are given.
Our universal relation is applied to several examples to show nontrivial constraints. In particular, a gapped ground state at a fractional charge filling per physical unit cell must have either a nonvanishing Hall conductivity or anyon excitations, excluding a trivial Mott insulator.
\end{abstract}

\keywords{Quantum Hall effects, Lieb-Schultz-Mattis theorem, Many-body polarization, Symmetry fractionalization}

\pacs{}

\maketitle

\tableofcontents




\section{Introduction}

Quantum Hall effects (QHE) of 2-dimensional electron gas (2DEG)~\cite{Klitzing1980,Tsui1982} exemplify the first class of topological phases~\cite{Wen2004B}, characterized by their quantized bulk response functions and gapless edge excitations.
In fact, QHE has been a source of many concepts that have become essential in more general quantum many-body problems.
One of the important directions in quantum many-body theory is to find a general constraint based on symmetries of the system. Such a constraint would be a guiding principle in classifying a wide variety of systems, and sometimes in designing systems with a desired property. QHE has been also instrumental in developing this type of approach. Laughlin's gauge argument for the quantization of the Hall conductance is a beautiful example, which inspired many other developments.

The Hall conductivity may also be constrained by other symmetries.
This is evident in the Hall effect of 2DEG in a free 2-dimensional space, where
the Hall conductivity $\sigma_{xy}$ is fully determined by filling factor $\nu$, the ratio of electron number density $\bar\rho$ and flux density $B/\Phi_0$:
\begin{eqnarray}\label{filling factor}
\sigma_{xy}^{\text{2DEG}}=\nu\frac{e^2}{h},~~~\nu=\frac{\bar\rho}{B/\Phi_0}=\frac{h\bar\rho}{eB}.
\end{eqnarray}
Guaranteed by Galilean invariance, this powerful relation remains valid in the presence of interactions, applying to both integer and fractional QHEs {in the free space}.

{
QHE on a periodic lattice is a more difficult problem.
Despite the complicated nature of the spectrum\cite{Hofstadter1976},
the Hall conductance is topologically
quantized\cite{TKNN,Kohmoto1985,Niu1985,Haldane1988}.
This observation led to the foundation of topological quantum many-body physics,
and has also received renewed interest
recently\cite{Sun2011,Tang2011,Neupert2011,Parameswaran2013a}.
The absence of the Galilean invariance in the lattice
implies the breakdown of the simple relation~\eqref{filling factor}.}
{Nevertheless, the periodic lattice has a discrete translation symmetry.
Therefore, we can ask a natural question:}
is there a relation similar to (\ref{filling factor}) for QHEs in a periodic crystal? More specifically, given the electron density and magnetic field, to what extent can we determine the Hall conductivity for QHEs on a discrete lattice?
In fact, the Lieb-Schultz-Mattis (LSM) theorem~\cite{Lieb1961} and its generalizations~\cite{OYA1997,Oshikawa2000,Hastings2005}, which are a filling-enforced constraint~\cite{PWZV2016} on quantum many-body systems, may be also be understood as a remnant of Galilean invariance on a discrete lattice.
Therefore it would be natural to expect a generalization of Eq.~\eqref{filling factor} to lattice systems.

In this paper, we demonstrate that this is indeed the case by deriving a universal relation between the quantized Hall conductivity and filling (particle number per unit cell), which generalizes Eq.~\eqref{filling factor}.
In quantum mechanics, a magnetic field must be represented in terms of vector potential. As a consequence, the original, physical translation symmetry is lost in the Hamiltonian in the presence of the magnetic field. This is the case even in the free space. Nevertheless, there is a remnant of the translation symmetry that is called a magnetic translation symmetry~\cite{Zak1964a,Zak1964}.
Laughlin's gauge argument and (generalized) LSM theorem are both consequences of large gauge invariance. The former constrains the Hall conductivity based on just the gauge invariance and the energy spectrum of the system. The latter constrains the energy spectrum of the system based on the ``filling'' or the particle density, in the presence of the discrete lattice translation symmetry. Our result is a combination of these two fundamental constraints in quantum many-body problem, relating the quantized Hall conductivity and the particle density. We will give the universal relation~\eqref{thm} first for the integer QHE case, and then later the corresponding result~\eqref{thm:fractional} for the fractional QHE. As is the case with the Laughlin's argument and the LSM theorem, our argument is very general and applies to a wide range of interacting systems of either bosons or fermions. {While some formulae corresponding to a part of our results
were reported earlier\cite{Dana1985,AvronYaffe1986,KolRead1993},
our perspective based on the LSM-type
theorems elucidates the deep physical meaning of the universal relations.
In fact, we will present several applications which demonstrate
their surprising power.}

The article is organized as follows. We set up the problem and summarize main results in Section~\ref{sec:setup}.
In Section~\ref{sec:flux},
as a preparation, the flux insertion process and the associated
momentum counting argument are introduced.
Then we prove the filling-enforced constraint on the quantized Hall conductivity for integer QHEs with a ``cut and glue'' argument, based on gapless edge states in Section~\ref{sec:cut and glue}.
In Section~\ref{sec:polarization}, we review the many-body polarization
as formulated by Resta and Sorella.
Then the many-body polarization is applied to derive the filling-enforced constraint for IQHEs without relying on the ``cut and glue'' procedure,
in Section~\ref{sec:IQHE}.
The filling-enforced constraint on the quantized Hall conductivity is then generalized for FQHEs in Section~\ref{sec:FQHE}.
An alternative perspective based on a fractionalization algebra
of magnetic translation symmetry is discussed in Section~\ref{sec:frac_mag_trans}.
In Section~\ref{sec:non-symmorphic}, generalization of our results to non-symmorphic lattices is discussed.
Applications of our filling-enforced constraints to various systems of interest are discussed in Section~\ref{sec:applications} before the concluding remarks in Section~\ref{sec:conclusion}.

\section{Setup and Main Results}

\label{sec:setup}

We consider a quantum many-particle system on
a periodic two-dimensional potential/lattice in
a background magnetic field, with the magnetic flux
\begin{equation}
 \phi = \frac{p}{q},
\label{eq.phi_pq}
\end{equation}
in the unit of flux quantum $\Phi_0 = h/e$ per unit cell,
where $p$ and $q$ are mutually prime integers.
In the following, we set $\hbar=e=1$ so that $\Phi_0 = 2\pi$.
We denote the primitive Bravais vectors of the lattice
as $\vec{a}_{1,2}$,
and corresponding translation operators as $T_{1,2}$.
The continuum version of the Hamiltonian reads
\begin{align}
 \calH & =
 \int d\vec{r} \;
 \frac{1}{2m}
 \psi^\dagger(\vec{r})
 \left[
\left(-\imth\vec{\nabla} - \vec{A}(\vec{r}) \right)^2
+ V(\vec{r})
\right]
 \psi(\vec{r}) \\
\notag
& +
 \int d\vec{r} \int d\vec{r'}\;
  U(\vec{r},\vec{r'}) \rho(\vec{r}) \rho(\vec{r'}),
\end{align}
where $\rho \equiv \psi^\dagger \psi$ is the particle number density,
and $V$ and $U$ are invariant under $T_{1,2}$.
It is also convenient to introduce the reciprocal vectors
$\vec{g}_{1,2}$ which satisfy
\begin{equation}
 \vec{a}_\alpha \cdot \vec{g}_\beta = \delta_{\alpha \beta} .
\end{equation}
While there is a degree of freedom in gauge choice, in this paper
we choose the Landau gauge
\begin{equation}
 \vec{A}(\vec{r}) = 2 \pi \phi r_1 \vec{g}_2,
\label{eq.LandauGauge}
\end{equation}
where
\begin{equation}
 r_\alpha = \vec{g}_\alpha \cdot \vec{r},
\label{eq.def.r_alpha}
\end{equation}
for $\alpha =1,2$.
Because of the position dependence of the vector potential,
the Hamiltonian is not invariant under the translation $T_1$.
Nevertheless, one can define a set of magnetic translation
operators $\tilde{T}_{1,2}$, which supplement the primitive
translations by appropriate gauge transformations and leave
the Hamiltonian invariant~\cite{Zak1964a,Zak1964}.
For the Landau gauge~\eqref{eq.LandauGauge},
we choose
\begin{align}
 \tilde{T}_1 &= \exp{\left( 2 \pi \imth \phi
 \int d\vec{r} \; r_2 \rho(\vec{r}) \right)} T_1,
\label{eq.T1}
\\
 \tilde{T}_2 &= T_2.
\label{eq.T2}
\end{align}
These magnetic translation operators satisfy the commutation relation
\begin{eqnarray}\label{mag trans}
\tilde T_1 \tilde T_2=
e^{2\pi\imth\phi\hat{N}}\tilde T_2\tilde T_1,
\end{eqnarray}
where
\begin{equation}
 \hat{N} = \int d\vec{r}~\hat\rho{(\vec{r})}
\end{equation}
is the total charge (in unit of the elementary charge $e$)
of the system.
This ``magnetic translation algebra'' may be regarded as
a defining feature of the system in the presence of the magnetic
field.
The magnetic translation algebra can be also defined for
a lattice model under a uniform magnetic field.
Our analysis in the following also applies to such lattice models
as well.

In particular, it is helpful to consider a system on the
square lattice with a uniform magnetic field, as a simplest
example, for the sake of illustrating the problem.
For the square lattice
with only the nearest-neighbor hoppings, the vector
potential on the lattice in the Landau gauge reads
\begin{equation}
\begin{aligned}
 A_1 (\vec{r}) & =  0,
\\
 A_2 (\vec{r}) & =  2\pi \phi r_1 ,
\end{aligned}
\label{eq.LandauGauge.sqlattice}
\end{equation}
where $\vec{r}=(r_1,r_2)^T \in \mathbb{Z}^2$ refers to the
location of the lattice site.
The vector potential enters the hopping term in the Hamiltonian as
\bea
&\notag - t_{\vec{r}+\vec{e}_1,\vec{r}} e^{\imth A_1(\vec{r})}
c^\dagger(\vec{r}+\vec{e}_1) c(\vec{r})\\
& - t_{\vec{r}+\vec{e}_2,\vec{r}} e^{\imth A_2(\vec{r})}
c^\dagger(\vec{r}+\vec{e}_2) c(\vec{r}) + \mbox{H.c.}
\eea
where $\vec{e}_1 = (1,0)^T$ and $\vec{e}_2 = (0,1)^T$.
The primitive translations $T_{1,2}$ are just the
lattice translations by $\vec{e}_{1,2}$.
The magnetic translation operators on the square lattice are then
\begin{align}
 \tilde{T}_1 &= \exp{\left( 2 \pi \imth \phi
 \sum_{\vec{r} \in \mathbb{Z}^2} \; r_2 n(\vec{r}) \right)} T_1,
\\
 \tilde{T}_2 &= T_2,
\end{align}
where $n(\vec{r})$ is the particle number operator at the
site $\vec{r}$.
They satisfy the same magnetic translaton algebra~\eqref{mag trans}.


The non-commutativity of magnetic translation operators
$\tilde{T}_1$ and $\tilde{T}_2$ prevents us from
using the common techniques such as Fourier transforming
to the momentum space.
Thus it is often convenient to use the commuting set of
magnetic translation operators, say $\left( \tilde{T}_1 \right)^q$
and $\tilde{T}_2$.
This corresponds to considering a ``magnetic unit cell'',
which is $q$ times larger than the original, physical unit cell.
Throughout this paper, ``unit cell'' (without ``magnetic'' in its
front) refers to the original unit cell and not to the
magnetic unit cell.

Now let us consider a system defined on a torus.
We consider a system consisting of $L_1 \times L_2$ unit cells,
namely $0 \leq r_\alpha < L_\alpha$, where
$r_\alpha = \vec{g}_\alpha \cdot \vec{r}$.
The uniformity of the magnetic field including on the boundaries
requires the generalized periodic boundary conditions
\begin{equation}
 \psi(\vec{r} ) = \left(\tilde{T}_\alpha\right)^{L_\alpha} \psi(\vec{r}) .
\end{equation}
For the Landau gauge~\eqref{eq.LandauGauge}, it is explicitly given as
\begin{align}
 \psi(\vec{r}) & = e^{2 \pi i \phi L_1 r_2}  \psi(\vec{r}+L_1 \vec{a}_1),
\label{eq.gpbc1}
\\
 \psi(\vec{r}) & = \psi(\vec{r}+L_2 \vec{a}_2).
\label{eq.gpbc2}
\end{align}
The consistency of the relation between $\psi(\vec{0})$
and $\psi(L_1 \vec{a}_1 + L_2 \vec{a}_2)$, which can be obtained
by applying $(\tilde{T}_1)^{L_1}$ and
$(\tilde{T}_2)^{L_2}$ in two different orders, requires
\begin{equation}
 \phi L_1 L_2 \in \mathbb{Z} .
\label{eq.total_flux_quantization}
\end{equation}
This is equivalent to the condition that the total magnetic
flux piercing the system is an integral multiple of
the flux quantum $\Phi_0$.
Therefore, $L_1 L_2$ must be an integral multiple of $q$.
Likewise, in a closed system, the total particle number
\begin{equation}
 N = \bar{\rho} L_1 L_2,
\end{equation}
where $\bar{\rho}$ is the average number of particles per unit cell,
must be an integer.
If
\begin{equation}
 \bar{\rho} = \frac{p'}{q'},
\label{eq.barrho_ppqp}
\end{equation}
where $p'$ and $q'$ are mutually coprimes,
$L_1 L_2$ must contain the factor $q'$ as well as $q$.

At this point, there is a freedom in assigning the factor $q$
to $L_1$ and $L_2$. It is possible to choose $L_1$ as an
integral multiple of $q$ and $L_2$ as a coprime with $q$,
or vice versa.
However, the choice of $L_1$ and $L_2$ is further restricted
by the requirement of (magnetic) translation symmetry.

In the Landau gauge~\eqref{eq.LandauGauge}, the magnetic
translation operator $\tilde{T}_2$ in the $2$-direction is
identical to the original lattice translation operator $T_2$, as
in Eq.~\eqref{eq.T2}.
It thus appears that the system is always invariant under
$\tilde{T}_2=T_2$.
However, for a generic choice of $L_1$, the generalized periodic
boundary condition~\eqref{eq.gpbc1} required for the torus
breaks the translation invariance under $T_2$.
For the system on the torus to be invariant under $\tilde{T}_2=T_2$,
the phase factor $e^{2\pi i \phi L_1 r_2}$ must be
invariant under $r_2 \to r_2 +1$
This requires $\phi L_1 \in \mathbb{Z}$, namely
\begin{equation}
 L_1 = q l_1,
\label{eq.L1_ql1}
\end{equation}
where $l_1$ is an integer.

On the other hand,the magnetic translation operator $\tilde{T}_1$,
Eq.~\eqref{eq.T1},
in the present Landau gauge can be written as
\begin{equation}
 \tilde{T}_1 = \left(\calU_2 \right)^{\phi L_2} T_1 ,
\label{eq.tT1_U2}
\end{equation}
where
\begin{equation}
 \calU_\alpha =  \exp{\left( \frac{2 \pi \imth}{L_\alpha}
 \int d\vec{r} \; r_\alpha\hat\rho(\vec{r}) \right)},
\label{eq.def.calU}
\end{equation}
for $\alpha=1,2$.
On the torus with the (generalized) boundary condition,
$\calU_2$ is nothing but the ``fundamental'' large gauge
transformation along the $2$-direction.
An integer power of $\calU_2$ is also a large gauge
transformation on the torus, as well.
However, a fractional power of $\calU_2$ is not a well-defined
large gauge transformation on the torus.
The Hamiltonian on the torus is invariant under the magnetic translation
$\tilde{T}_1$ only if $\left(\calU_2\right)^{\phi L_2}$ is a large
gauge transformation, namely only if
\begin{equation}
 \phi L_2 \in \mathbb{Z} .
\label{eq.phi_L2}
\end{equation}
This requires $L_2$ to be an integral multiple of $q$.
In this sense, with the Landau gauge~\eqref{eq.LandauGauge},
the Hamiltonian on the torus has the magnetic translation invariance
under $\tilde{T}_1$, only when $L_2$ is an integral multiple of $q$.
We note that, even when $L_2$ is not an integral multiple of $q$,
the system on the torus of
the size $L_1 \times L_2$ is also perfectly well-defined
with uniform flux with the generalized periodic boundary conditions
\eqref{eq.gpbc1} and~\eqref{eq.gpbc2}.



The major results of this paper are summarized in the following
theorems.

\begin{theorem} {\bf Filling-enforced constraint on $\sigma_{xy}$ for IQHE:}\\
In a generic two-dimensional (2d) system of bosons and/or fermions
preserving magnetic translation symmetry (\ref{mag trans}) and $U(1)$ charge conservation, if it has a
unique gapped ground state on torus, its Hall conductivity
must satisfy the following condition:
\begin{eqnarray}
{\tilde{\sigma}_{xy}}\cdot{\phi}={\bar{\rho}}\mod1,
\label{thm}
\end{eqnarray}
where $\tilde{\sigma}_{xy}= \sigma_{xy}\cdot h/e^2$ is the
Hall conductivity in the unit of $e^2/h$, and
$\bar{\rho}$ is the number of particles (or the charge in unit of fundamental charge $e$) per unit cell.
\label{theorem.IQHE}
\end{theorem}
This theorem is the lattice analog of (\ref{filling factor}) for 2DEG in
continuum.



Furthermore, it can be extended to Fractional Quantum Hall effects (FQHEs)
with topologically degenerate ground states, as follows:
\begin{theorem} {\bf Filling-enforced constraint on $\sigma_{xy}$ for FQHE:}\\
In a generic gapped two-dimensional (2d) system of bosons and/or fermions preserving magnetic translation symmetry (\ref{mag trans}) and $U(1)$ charge conservation, its Hall conductivity
$\tilde{\sigma}_{xy}$ in unit of $e^2/h$ must satisfy the following condition
\begin{eqnarray}\label{thm:fractional}
\bar{\rho} =\tilde{\sigma}_{xy}\cdot\phi+\frac{\theta_{F,a}}{2\pi}\mod1.
\end{eqnarray}
where $\theta_{F,a}$ is mutual statistical angle between fluxon ($F$) and background anyon ($a$)~\cite{Essin2013,Barkeshli2014} in each unit cell.
When there are $n$ degenerate ground states,
\begin{equation}
 n \frac{\theta_{F,a}}{2\pi} \in \mathbb{Z} .
\label{eq.thetaFa_gsd}
\end{equation}
\label{theorem.FQHE}
\end{theorem}

{The non-interacting fermion version
of Theorem~\ref{theorem.IQHE} is known as
the Diophantine equation~\cite{Dana1985}.
It was generalized to many-body systems
by Avron and Yaffe~\cite{AvronYaffe1986}
and studied further in Ref.~\cite{KolRead1993}.
The many-body version of the Diophantine equation
in Ref.~\cite{AvronYaffe1986} corresponds to
the above theorems at a fixed system size and
number of particles, with a given number of degenerate ground states.
However, we believe that the theorem stated as above in terms of
$\phi$ and $\bar{\rho}$ is useful in many ways.
For example, it reveals a deep connection to the LSM-type theorems,
leading to natural generalizations such as
Theorem~\ref{theorem.FQHE.ns}
for non-symmorphic lattices in Section~\ref{sec:non-symmorphic}.
Furthermore, it gives rise to powerful applications,
in particular the one showing the stability of gapless phases in $\pi$-flux
systems, as we will discuss in Section~\ref{sec:applications}.
The relation to the statistical angle $\theta_{F,a}$ is also
a novel perspective introduced in this paper.}

Clearly, Theorem~\ref{theorem.IQHE}
is simply a special case without bulk anyons, of the
latter more general Theorem~\ref{theorem.FQHE}.
When $\phi=0$ magnetic translation (\ref{mag
trans}) reduces to usual lattice translations, and
(\ref{thm:fractional}) reduces to the well-known fact that a gapped 2d
phase with fractional filling per unit cell necessarily leads to
topological order if translations $T_{1,2}$ are
preserved~\cite{Hastings2004,Hastings2005,Oshikawa2006,Cheng2016}. As will be shown later, because the magnetic translation symmetry determines only the fractional part of $\phi$, the background anyon $a$ per unit cell has the following ambiguity: When
we increase the flux density $\phi$ by $\Delta\phi=1$, there is one
extra ``anti-fluxon'' per unit cell:
\bea\label{redef bg flux+anyon}
\phi\rightarrow \phi+1,~~~a\rightarrow a-F
\eea
It then follows that $\theta_{F.a} \to \theta_{F,a} - \theta_{F,F}$
under this transformation.
The consistency of the Theorem~(\ref{thm:fractional}) with this
transformation leads to the identity
\begin{equation}
 e^{\imth\theta_{F,F}}=e^{2\pi\imth\tilde \sigma_{xy}}.
\label{eq.sigmaxy_FF}
\end{equation}
This is actuall a well known result~\cite{Cheng2016}
that Hall conductivity equals
the mutual statistical angle of two fluxons in unit of $2\pi$.

The theorem (\ref{thm:fractional}) can be proved for both Abelian and
non-Abelian topological orders, under the assumption of symmetry
fractionalization (i.e. we are limited to the situations that magnetic
translation symmetry does not change anyon types). Note that even in
non-Abelian cases, background anyon $a$ per unit cell must be an Abelian
anyon to be compatible to translational symmetry operation in
topological orders~\cite{Barkeshli2014,Cheng2016}. Meanwhile, the fluxon
must also obey Abelian statistics even in non-Abelian FQH
states~\cite{Cheng2016}.

%

\section{Flux insertion}
\label{sec:flux}

\subsection{Large gauge invariance and momentum counting}
\label{sec:mom_count}

Suppose that the system is in a ground state initially
at $t=0$.
Let us define the ``flux insertion operators'' $\calF_\alpha(\Phi)$,
where $\alpha=1,2$, in the following way.
They represent the time evolution operators for a process of
adiabatic
insertion of flux $\Phi$ through the ``hole'' encircled by
a loop in $\alpha$ direction, where the periodic boundary
condition is imposed.
It should be noted that, this flux is an Aharonov-Bohm flux
which does not directly touch the particles on the two-dimensional system,
and should be distinguished from the magnetic flux
($\phi \Phi_0$ per unit cell) which is piercing through the system
(and ``perpendicular'' to the two-dimensional plane).

For example, we choose the time evolution of the Hamiltonian as
\begin{equation}
 \calH_\alpha(\Phi,t) =
 \calH_\alpha(A_\alpha = \frac{\Phi}{L_\alpha}\frac{t}{T}),
\end{equation}
for a large enough time $T$.
Then
\begin{equation}
 \calF_\alpha(\Phi) \equiv \calT \exp{\left(
- i \int_0^T dt \; \calH_\alpha(\Phi,t) \right)} ,
\end{equation}
where $\calT$ denotes the time ordering.
After the flux insertion process $\calF_\alpha(\Phi)$,
the Hamiltonian becomes $\calH(\Phi)$ which contains the flux $\Phi$.

On the other hand, the large gauge transformation is
characterized by the identity
\begin{equation}
 \calH_\alpha(\Phi_0) =
  \calU_\alpha  \calH_\alpha(0) \left( \calU_\alpha \right)^{-1} .
\label{eq.UHU}
\end{equation}
Thus, if we consider the insertion of unit flux quantum $\Phi_0$ by
$\calF_\alpha(\Phi_0)$, the inserted flux can be eliminated precisely
by the large gauge transformation~\eqref{eq.def.calU}.
It is thus convenient to consider the combined operation
\begin{equation}
 \tilde{\calF}_\alpha(\Phi_0)
  \equiv \left(\calU_\alpha \right)^{-1} \calF_\alpha(\Phi_0) ,
\label{eq.def.tildecalF}
\end{equation}
of the insertion of the unit flux quantum $\Phi_0$ and the
large gauge transformation.
This is the operator what has been used in
various contexts~\cite{Oshikawa2000,Oshikawa-Lutt2000,OshikawaSenthil2006}.
After this process,
we get back to the original Hamiltonian $\calH_\alpha(0)$.
However, in general, the final state
$\tilde{\calF}_\alpha(\Phi_0)|\Psi_I\rangle$ after
the evolution is different from
the initial state $|\Psi_I\rangle$.

To see this, it is useful to invoke the ``momentum counting''
argument~\cite{Lieb1961,OYA1997,Oshikawa2000}.
As in Sec.~\ref{sec:setup},
let us consider the system consists of $L_1 \times L_2$ unit cells.
In general, we have
\begin{equation}
T_\alpha \calU_\alpha {T_\alpha}^{-1}\calU_\alpha^{-1}
=e^{-\frac{2\pi\imth}{L_\alpha} \int d\vec{r} \; \rho{(\vec{r})}}
=e^{-{2\pi\imth\bar{\rho} L_{\bar{\alpha}}}},
\label{eq.T_calU}
\end{equation}
where $\bar{1}=2$ and $\bar{2}=1$.
In our setup with the Landau gauge~\eqref{eq.LandauGauge},
$\tilde{T}_2 = T_2$ commutes with the Hamiltonian, when
the periodic boundary condition is imposed in $2$-direction.
Thus the total momentum $P_2$ in $2$-direction defined by
\begin{equation}
 T_2 = e^{i P_2}
\end{equation}
is a good quantum number, modulo $2\pi$.

On the other hand, since the introduction of the uniform vector
potential corresponding to flux insertion in the hole
does not break the translation symmetry,
\begin{equation}
 [ T_2, \calF_2(\Phi)] = 0 .
\end{equation}
Combining this and Eq.~\eqref{eq.T_calU} for $\alpha=2$
with the definition~\eqref{eq.def.tildecalF},
we find
\begin{equation}
 T_2 \tilde{\calF}_2(\Phi_0) = \tilde{\calF}_2(\Phi_0) T_2
e^{2\pi\imth\bar{\rho} L_1}.
\label{eq.T2_tildecalF}
\end{equation}
Eq.~\eqref{eq.T2_tildecalF} implies that the momentum $P_2$
of the final state differs from that of the initial state by
\begin{equation}
 \Delta P_2 \equiv 2\pi \bar{\rho} L_1 \mod{2\pi} .
\label{eq.DeltaP2}
\end{equation}
If $\bar{\rho} L_1$ is not an integer,
the final state
$\tilde{\calF}_\alpha(\Phi_0)|\Psi_I\rangle$ must be orthogonal to
the initial state $|\Psi_I\rangle$.

We note that, up to this point, we do not need an assumption
of adiabaticity of the flux insertion process.
The present result is valid however fast the flux is inserted,
and whether the system has an excitation gap or not.
Moreover, we assumed the periodic boundary condition in the $2$-direction,
but not in the $1$-direction at this point.
Thus the momentum counting~\eqref{eq.DeltaP2} is also valid
when the open boundary condition is imposed in the $1$-direction,
for example.

In the application to QHE, a system with open boundaries generally
has gapless edge states, even when there is a non-vanishing excitation
gap in the bulk.
The flux insertion/momentum counting argument can still be useful
in the presence of the gapless edge states,
as we will discuss later in Section~\ref{sec:cut and glue}.

\subsection{Momentum counting on a torus}
\label{sec:mom_count_torus}

In fact, it is often convenient to consider gapped system
without edge states.
For this, we impose the periodic boundary conditions in both
directions, or equivalently consider the system on a torus
of the size $L_1 \times L_2$.
This introduces several subtleties.
First, to make the system well-defined on the torus,
the total magnetic flux piercing the system must be an integral
multiple of the unit flux quantum,
as in Eq.~\eqref{eq.total_flux_quantization}.
It follows that $L_1 L_2$ must be an integral multiple of $q$.

As we have discussed in Section~\ref{sec:setup}, there is some freedom
in choosing $L_1$ and $L_2$.
Since the translation invariance under $T_2$ is essential for
the momentum counting argument,
on the torus we have to choose $L_1$ to be
an integral multiple of $q$, as in Eq.~\eqref{eq.L1_ql1}.
Under this condition, Eq.~\eqref{eq.T_calU} for $\alpha=2$ reads
\begin{equation}
T_2 \calU_2 {T_2}^{-1}\calU_2^{-1}
=e^{-\frac{2\pi\imth}{L_2} \int d\vec{r} \; \rho{(\vec{r})}}
=e^{-{2\pi\imth\bar{\rho} q l_1}},
\label{eq.T_calU.2}
\end{equation}
and thus the momentum gain due to the flux insertion
(and large gauge transformation) $\tilde{\calF}_2(\Phi_0)$
is
\begin{equation}
 \Delta P_2 \equiv 2\pi \bar{\rho} L_1 \mod{2\pi}
\equiv 2\pi \bar{\rho} q l_1 \mod{2\pi}  .
\end{equation}
$l_1$ can be any integer, and we choose it as a coprime with $q'$.

In the presence of the gap, we may require
flux insertion process $\calF$ to be adiabatic,
assuming that the gap never closes during the
flux insertion~\cite{Oshikawa2000}.
Under this condition, if the initial state $|\Psi_I\rangle$
is taken as a ground state $|\Psi_0 \rangle$, which
is also an eigenstate of $T_2$, the final state
$\tilde{\calF}_2(\Phi_0) | \Psi_0 \rangle$
must also be a ground state.
Then, if $\bar{\rho} q$ is not an integer, there must be
multiple degenerate ground states, or gapless excitations.
This is the generalized LSM
theorem applied to a system under uniform magnetic field
($\phi=p/q$ flux per unit cell).
The statements of the standard LSM theorem apply, by replacing
$\bar{\rho}$ by $\bar{\rho} q$.
Namely, under the magnetic field,
the particle number per \emph{magnetic unit cell}, which is
$q$ times larger than the original unit cell, is the relevant
parameter for the generalized LSM theorem.
In particular, if we assume a unique ground state with a gap on the torus,
\begin{equation}
 \bar{\rho} q \in \mathbb{Z}\Leftrightarrow\bar\rho\phi=0\mod1 ,
\label{eq.rho_q_int}
\end{equation}
or equivalently $q/q' \in \mathbb{Z}$.
This corresponds to the integer number of particles per
magnetic unit cell.
In the limit of non-interacting electrons, Eq.~\eqref{eq.rho_q_int}
allows the Fermi level to lie in a band gap between
bands defined on the magnetic Brillouin zone.
However Eq.~\eqref{eq.rho_q_int} holds in more general interacting
systems, as long as the system is gapped
with the unique ground state on a torus.

At this point, the LSM theorem gives no constraint
when $\bar{\rho} q \in \mathbb{Z}$, even if $\bar{\rho}$ is fractional.
Nevertheless, in the following, we will go beyond the LSM theorem
and show that a different type of general constraint arises
even when $\bar{\rho} q \in \mathbb{Z}$.

\subsection{Flux insertion and effective symmetries}
\label{sec:eff_sym}

We have seen that the adiabatic insertion of the unit flux quantum
$\calF_\alpha(\Phi_0)$ and the fundamental large gauge transformation
$\calU_\alpha$ have a similar effect of introducing the
unit flux quantum $\Phi_0$ in the ``hole'' of the torus.
However, they are quite different operators
(and thus $\tilde{\calF}_\alpha(\Phi_0) \neq 1$).
In fact, $\calF_\alpha$ correspond to application of an
electric field in $\alpha$-direction, while $\calU_\alpha$
does not.
Physically, the applied electric field can accelerate particles.
Thus, the flux insertion process generally changes
the energy of a given initial state.
As a consequence, the analog of Eq.~\eqref{eq.UHU},
generally does \emph{not} hold for $\calF_\alpha(\Phi)$,
even for $\Phi=\Phi_0$.
Nevertheless, when the system is gapped, the adiabaticity of $\calF_\alpha$
implies that
\begin{equation}
\calH_\alpha(\Phi_0) \sim
  \calF_\alpha(\Phi_0)  \calH_\alpha(0)
\left( \calF_\alpha(\Phi_0) \right)^{-1},
\end{equation}
holds, when \emph{restricted onto the ground-state subspace} of the
entire Hilbert space.
(This holds because the initial and final states are both
ground state(s) of the physically equivalent Hamiltonian and
thus have the same eigenvalue.)
It also follows that
\begin{equation}
 [ \tilde{\calF}_\alpha(\Phi_0), \calH(0)] \sim 0,
\label{eq.tildecalF_calH}
\end{equation}
\emph{in the ground-state subspace}.
In this sense, $\tilde{\calF}_\alpha(\Phi_0)$ effectively acts
as a symmetry generator, as far as the ground states are concerned.

Furthermore, one can certainly consider an adiabatic flux insertion,
$\calF_\alpha(\Phi)$, for any $\Phi$ which is not necessarily
an integer multiple of the unit flux quantum $\Phi_0$.
Let us consider a finite-size system of $L_1 \times L_2$
unit cells on a torus with the flux~\eqref{eq.phi_pq}
per unit cell, in the Landau gauge~\eqref{eq.LandauGauge}
(or its lattice version).
Let us choose $L_1$ to be an integral multiple of $q$,
as in Eq.~\eqref{eq.L1_ql1},
so that the system retains the translation symmetry $T_2$,

In order to derive strongest possible restrictions,
we choose $L_2$ to be a coprime with $q$.
As we remarked in Sec.~\ref{sec:setup},
this choice is allowed as $L_1$ is already
chosen to be an integral multiple of $q$.
A drawback of this choice of $L_2$ is that,
the magnetic translation operator $\tilde{T}_1$
is in fact not well-defined because $\phi L_2$ is not an integer,
as we have discussed in Eq.~\eqref{eq.tT1_U2}.
Nevertheless, here we can use the adiabatic flux insertion,
instead of the large gauge transformation, to define
an analogue of the magnetic translation operator.
The adiabatic flux insertion $\calF_2(\phi L_2 \Phi_0)$ exactly
compensates the change in the vector potential induced
by the primitive translation $T_1$.
Therefore,
\begin{equation}
 \tilde{\tilde{T}}_1 \equiv \calF_2(\phi L_2 \Phi_0) T_1
\label{eq.tiltilT1}
\end{equation}
maps a ground state of the original Hamiltonian $\calH(0)$ to
a ground state of the same Hamiltonian, and thus
can be regarded as another ``effective symmetry generator''
in the same sense as Eq.~\eqref{eq.tildecalF_calH}.
This may be regarded as a remnant of the magnetic translation
symmetry on the torus, although the magnetic translation operator
$\tilde{T}_1$ in the standard sense is ill-defined in this
setup.

As discussed earlier, $\phi L_2$ not being an integer does not
prevent adiabatically inserting the ``fractional flux''
$\phi L_2 \Phi_0$, while the large gauge transformation
is ill-defined for a fractional flux.
From the commutation relations~\eqref{eq.calF2_calU1} and
\eqref{eq.T_calU}, we find
\begin{equation}
 \tilde{\tilde{T}}_1 \calU_1 = \calU_1 \tilde{\tilde{T}}_1
   e^{2\pi\imth L_2 ( \phi \tilde{\sigma}_{xy} - \bar{\rho})} .
\label{eq.tiltilT1_calU1}
\end{equation}
This operator $\tilde{\tilde{T}}_1$ will prove useful,
as we will demonstrate later in Section~\ref{sec:IQHE_periodic}.


\section{Filling-enforced constraint on $\sigma_{xy}$:
 a ``cut and glue'' proof using edge states}
\label{sec:cut and glue}

In the following we give a heuristic proof of the main theorem~(\ref{thm})
for integer QHEs.
This is based on a ``cut and glue'' procedure,
by examining how the gapless edge states which appears upon
cutting the system, respond to flux insertion.
It can be straightforwardly generalized to
non-symmorphic lattices, leading to a stronger condition
(\ref{thm:nonsymmorphic:fractional}).


We first consider the system on a torus of size $L_1 \times L_2$, as
discussed in the previous section.  To represent the uniform magnetic
field, we adopt the Landau gauge~\eqref{eq.LandauGauge}.  Here
we choose $L_1 = q l_1$, $l_1 =n_1 q +1$, and $L_2 = n_2 q$ with
$n_{1,2} \in \mbz$.  With this choice, we satisfy the
requirement~\eqref{eq.total_flux_quantization}.  Furthermore, the
magnetic translation operators $\tilde{T}_{1,2}$ are well-defined
and commute with the Hamiltonian, as
both conditions~\eqref{eq.L1_ql1} and \eqref{eq.phi_L2} are satisfied.

\begin{figure}
\includegraphics[width=1\columnwidth]{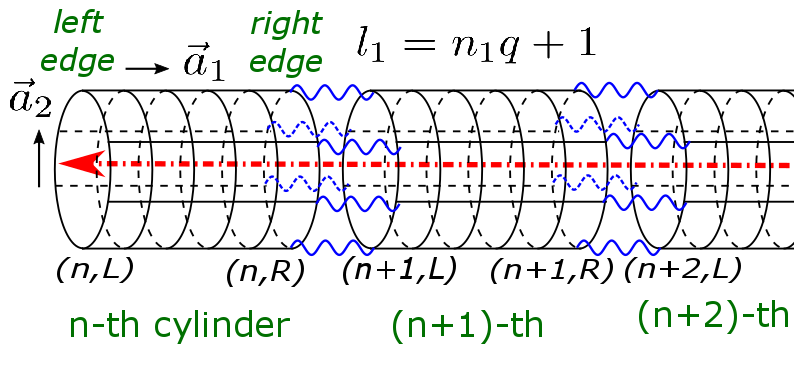}
\caption{(color online) An illustration of the flux insertion and cut-and-glue procedure. Red arrow of dotted dash line denotes inserting $2\pi$ flux adiabatically through the torus, }
\label{fig:torus glue}
\end{figure}

As illustrated in FIG. \ref{fig:torus glue}, this torus can be cut into
$q$ cylinders of the size $l_1 \times L_2$, where $L_2 = n_2 q$.  We
choose $l_1$ to be mutually prime with $q^\prime$ and $n_2=0\mod
q^\prime$ so that each cylinder can have an integer total number of
elementary particles, while keeping the fractional number
$\bar{\rho}=p^\prime/q^\prime$ of particle per unit cell.  While this is
not the only possible choice of the system size, it turns out to be
convenient as we will see in the following.  When $n_{1,2}$ are large
enough, each cylinder can be treated as a macroscopic system.
We emphasize that, although these $q$ cylinders are physically identical
to each other, the corresponding Hamiltonians are not identical.  This
can be seen by going back to the torus before it was cut into cylinders
and recalling the Landau gauge~\eqref{eq.LandauGauge} or its lattice
version such as Eq.~\eqref{eq.LandauGauge.sqlattice}.  The Hamiltonians
for two different cylinders are related by the magnetic translation
operator $\left(\tilde{T}_1 \right)^{l_1}$ or its integer powers, not
the simple translation.

In this Section, we assume that the system on the torus has a
non-vanishing gap and has a unique ground state.  This means that here
we consider an Integer Quantum Hall state, since a Fractional Quantum
Hall state must have topologically degenerate ground states on a
torus~\cite{TaoWu1984}.  We will discuss Fractional Quantum Hall
states later.

First we adiabatically insert a unit flux quantum $\Phi_0$ through each
cylinder (red arrow in FIG. \ref{fig:torus glue}).
Although each cylinder which appears after the cut does not have
a translation invariance in the $1$-direction, we can still apply
the momentum counting argument in the $2$-direction as discussed
in Section~\ref{sec:mom_count}.
As a result, for each cylinder,
we find that the momentum in the $2$-direction
acquires the shift~\eqref{eq.DeltaP2} where $L_1$ is replaced
by the width of the cylinder $l_1$, namely
\begin{equation}
 \Delta P_2 = 2 \pi \bar{\rho} l_1 .
\label{eq.DeltaP2_cyl}
\end{equation}
Because of our choice of $l_1$ so that it is a coprime with $q'$,
this is nontrivial modulo $2\pi$ and thus
the initial state and the final state after
the flux insertion $\tilde{\calF}(\Phi_0)$
must be orthogonal.
Following the logic of Refs.~\cite{Lieb1961,Oshikawa2000},
this leaves us two possibilities:
(i) \textit{there are (at least) two (quasi-)degenerate ground states below
the gap}, or
(ii) \textit{there are gapless excitations}.

\begin{figure}
\includegraphics[width=0.6\columnwidth]{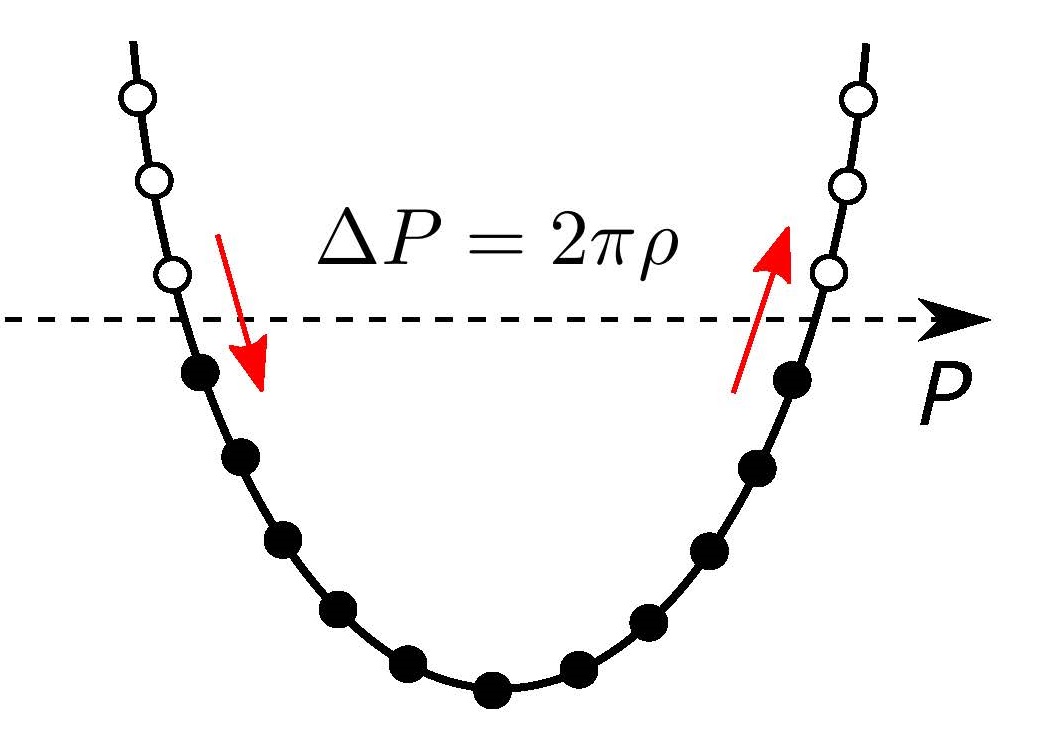}
\caption{(color online) An intuitive picture for generalized Luttinger's theorem (\ref{luttinger}).}
\label{fig:luttinger}
\end{figure}

Our assumption that the system has a non-vanishing gap with a unique
ground state on the torus then implies the presence of
gapless edge states on the cylinder.
Thus, under the assumption of a unique gapped ground state on a torus,
fractional filling with respect to the original, physical unit cell
$\bar\rho\not\in\mbz$ necessarily implies gapless
edge states in each cylinder.  Upon adiabatic insertion of one flux
quantum, a momentum shift~\eqref{eq.DeltaP2_cyl} is acquired
via level crossings on the two edges of each cylinder. By denoting the
momentum transfer through the left(right) edge of $n$-th cylinder (see
FIG. \ref{fig:torus glue}) as $\Delta P^{(n)}_{2,L(R)}$, we have
\begin{eqnarray}\label{flux insertion:1}
&\Delta P^{(n)}_{2,L}+\Delta P^{(n)}_{2,R}\equiv\Delta P_2=2\pi\bar\rho l_1\mod2\pi,\\
&\notag n=1,2,\cdots,q.
\end{eqnarray}
As illustrated in FIG. \ref{fig:luttinger}, the momentum transfer for each edge is always proportional to charge density $\rho_{n,L/R}$ (in unit of $e$) on each edge:
\begin{eqnarray}\label{luttinger}
\Delta P^{(n)}_{2,L(R)}=2\pi\rho_{n,L(R)},~~~\forall~n.
\end{eqnarray}
This can be viewed as a generalization of Luttinger's theorem to interacting systems. In Appendix \ref{app:chiral boson}, we proved this generalized Luttinger's theorem in the chiral boson description of edge states where the gapped bulk is captured by an Abelian Chern-Simons theory\cite{Wen1995}.

On the other hand, the $(n+1)$-th cylinder is related to the $n$-th
cylinder by magnetic translation $(\tilde T_1)^{l_1}$, which is a combination of pure translation $T_1^{l_1}$ and large gauge transformation $\calU_2^{\phi l_1L_2}$.


Therefore the $(n+1)$-th cylinder can be viewed as translating $n$-th
cylinder by $l_1 = n_1q+1$ unit cells along $\vec a_1$ direction, and
then
applying large gauge transformation $(\calU_2)^{\phi l_1L_2}= (\calU_2)^{p n_2 l_1}$ flux quanta through it. Due to the well-known chiral anomaly~\cite{Fujikawa2014B} in 1+1-D, the chiral edge states of QHEs is not invariant under large gauge transformations~\cite{Wen1995}. As shown in Appendix \ref{app:chiral boson}, a large gauge transformation $(\calU_2)^{\phi l_1L_2}$ can change the charge density of edge states by $\mp\phi l_1\sigma_{xy}$ for left(right) edges, where $\sigma_{xy}\in\mbz$ is the integer Hall conductance of the bulk. This leads to the following relation on
the charge density on the edges:
\begin{eqnarray}
&\notag\rho_{n+1,L}=\rho_{n,L}-\sigma_{xy}\phi l_1,\\
&\rho_{n+1,R}=\rho_{n,R}+\sigma_{xy}\phi l_1.\label{translating
cylinders}
\end{eqnarray}
Now let's glue the right edge of $n$-th and the left edge of $(n+1)$-th
cylinder together preserving translation $T_2$, and the entity of these
two cylinders becomes a larger cylinder of size $2l_1\times L_2$. With a
bulk excitation gap, such a gluing procedure won't change the charge
density beyond a finite correlation length and hence the charge
densities on two edges of the large cylinder are $\rho_{n,L}$ and
$\rho_{n+1,R}$. Repeating the flux insertion argument for this large
cylinder yields
\begin{eqnarray}
2\pi(\rho_{n,L}+\rho_{n+1,R})=2\Delta P_2=4\pi\bar\rho l_1\mod2\pi.
\label{flux insertion:2}
\end{eqnarray}
Combining (\ref{flux insertion:2}) with (\ref{flux
insertion:1})-(\ref{translating cylinders}) we obtain
\begin{eqnarray}
&\notag2\pi\sigma_{xy}\phi l_1=2\pi\bar\rho l_1\mod2\pi
\end{eqnarray}
A unique gapped ground state implies an integer $\sigma_{xy}\in\mbz$
without fractionally charged quasiparticles. By choosing $l_1=n_1q+1$
and $n_1=0\mod q^\prime$ for rational flux $\phi=p/q$ and commensurate
filling $\bar\rho=p^\prime/q^\prime$, clearly this leads to theorem
(\ref{thm}).

\section{Many-Body Electric Polarization}
\label{sec:polarization}

\subsection{Defining the many-body polarization as a bulk quantity}

The argument in the previous section suggests the universal
constraint based on the transport of electrons between the edge states.
However, there is a subtlety in the ``cut and glue'' procedure,
namely in the relation between the torus without cuts and
the cylinders obtained with the cuts.
In addition, the existence of the
gapless edge state makes the argument somewhat tricky.
However, when the system is gapless, the flux insertion process
cannot be adiabatic in the strict sense.
Although the momentum counting part of the argument does not
actually depend on the adiabaticity, it is difficult to control
the final state in the absence of the gap.

In fact, the same problem exists in the celebrated Laughlin's
gauge invariance argument for the quantization of Hall conductance,
as it depends on a cylinder geometry and accompanying gapless edge states.
While it may be still possible to define an adiabatic process
in a finite-size system, where generically there is a non-vanishing gap,
it is theoretically desirable to invoke an adiabatic argument
for a gapped system.
For this, it is advantageous to close the argument entirely within
the system with periodic boundary conditions, namely on a torus.
With the periodic boundary conditions, there is no gapless edge states.
The lack of edge states then appears to imply
that we cannot define the charge transport.
Fortunately, this is not quite the case.
In fact, there have been developments over several decades on
how to define the electric polarization with periodic boundary conditions.
This led to a beautiful formulation of electric polarization of
free electron systems in terms of geometric phases.
The theory is further generalized to interacting many-body systems.
What we would like to point out here is that the theory
of polarization in many-body systems is closely related to
large gauge invariance, and that we can derive useful identities
based on the relation.

Polarization vector $\vec{P}$ is roughly defined as
\emph{electric dipole moment per unit volume}.
On average, thus,
\begin{equation}
\bar{\vec{P}}
\sim \frac{1}{V} \left( \int d\vec{r} \;  \vec{r} \rho(\vec{r}) \right)
+ \vec{P}_0 ,
\label{eq.def.P}
\end{equation}
where $\vec{P}_0$ is the contribution from the immobile background
(``ions'') and $V$ is the volume of the system.
If the system is periodic with Bravais vectors $\vec{a}_\alpha$,
the bulk of the system is invariant under the translation
$\vec{r} \to \vec{r}+\vec{a}_\alpha$ ($\alpha=1,2$).
With the naive definition of the polarization~\eqref{eq.def.P},
however, it is not invariant under the translation.
This means that the polarization vector has some ambiguity
in the presence of periodic boundary conditions.
On the other hand, the change of $\vec{P}$ during a given process
should be uniquely defined.
In fact,
\begin{equation}
 \vec{j} = \frac{\partial \vec{P}}{\partial t}
\end{equation}
is the current density, which is a well-defined physical
quantity even with the periodic boundary conditions.
This suggests that it is possible to study polarization
even with the periodic boundary conditions.
In fact this has been a subject of intense studies
over several decades~\cite{RestaRevModPhys1994,RestaJPCMrev2002}.
Here we summarize the theory of polarization in many-body
interacting systems with periodic boundary conditions
developed by Resta and Sorella~\cite{RestaSorella1999}.
Furthermore, we point out its relation to the large
gauge invariance, and discuss the consequences.






As in the previous Section, let us consider a system consisting
of $L_1 \times L_2$ unit cells on a torus.
We define the ``total polarization'' in $\alpha$-direction as
\begin{equation}
 \calP_\alpha \equiv
\frac{1}{L_\alpha} \cdot
\left(
\int d\vec{r} \;  \cdot r_\alpha \rho(\vec{r})
\right),
\label{eq.def.totalP}
\end{equation}
where $r_\alpha$ is defined in Eq.~\eqref{eq.def.r_alpha}.
Comparing with Eq.~\eqref{eq.def.P}, $\bar{\vec{P}}$ and $\vec{\calP}$
are related as
\begin{equation}
 \calP_\alpha =
L_{\bar{\alpha}} \vec{g}_\alpha \cdot \bar{\vec{P}} + \mbox{const.},
\end{equation}
for $\alpha=1,2$.
This is naturally related to the total current
flowing in each direction:
\begin{equation}
I_\alpha = e \frac{\partial \calP_\alpha}{\partial t} .
\label{eq.I_calP}
\end{equation}


One can easily see that $\calP_1$ is invariant under $T_2$.
On the other hand, under $T_1$, we find the change in
\begin{equation}
 \Delta \calP_1 = [T_1, \calP_1] =
\frac{1}{L_1}
\int d\vec{r} \;  \vec{g}_1 \cdot \vec{a}_1 \rho(\vec{r})
=
\frac{N}{L_1} = \bar{\rho} L_2,
\end{equation}
where $\vec{r}$ refers to the lattice sites and
$n_{\vec{r}}$ is the particle number operator at the site $\vec{r}$.
Therefore, the total polarization $\calP_1$
would be only well-defined modulo $\bar{\rho}L_2 = p' L_2/q'$.


Resta and Sorella~\cite{RestaSorella1999} discussed the electric
polarization in general interacting many-body systems.  They introduced
an exponential of the total polarization operator~\eqref{eq.def.totalP}
as
\begin{equation}
  \calU_\alpha = e^{2\pi i \calP_\alpha} .
\label{eq.U_by_RS}
\end{equation}
They considered its ground-state
expectation value
\begin{equation}
 z_\alpha \equiv \langle \calU_\alpha \rangle, \label{eq.z}
\end{equation}
and then argued that the total polarization $\bar{\calP}_\alpha$ of the
ground state can be defined as
\begin{equation}
   z_\alpha = \left| z_\alpha \right| e^{2 \pi i \bar{\calP}_\alpha} .
\label{eq.P_from_z}
\end{equation}
This would determine $\bar{\calP}_\alpha$ modulo integer.
$\bar{\calP}_\alpha$ would be ill-defined if the expectation value
$z_\alpha$ vanishes.  Resta and Sorella argued that the vanishing of
$z_\alpha$ (in the thermodynamic limit) implies that the system is
conductor in which the polarization is ill-defined.
According to them, on the other hand,
$|z_\alpha|$ should approach $1$ in insulators so that
the polarization $\calP_\alpha$ has vanishing fluctuations
and takes the definite value $\bar{\calP}_\alpha$.
(See also Refs.~\cite{SWM2000,Resta_JPCM2002})
In this paper we assume that this statement
holds in insulators including Quantum Hall states.


In fact, the operator~\eqref{eq.U_by_RS}
is nothing but the fundamental large gauge
transformation operator (or equivalently
the LSM twist operator))~\eqref{eq.def.calU}, as pointed out in
Ref.~\cite{NakamuraVoit_2002}.
However, so far the large gauge invariance has not been exploited
in the context of the Resta-Sorella theory of many-body polarization.
The large gauge invariance based on $\calU_\alpha$
has been quite useful in deriving various universal constraints
on quantum many-body problem.
It is one of the goals of the present paper to
develop new applications of the large gauge invariance
in the context of the many-body polarization.
In later Sections, we shall demonstrate that
it is indeed fruitful.

\subsection{Many-body polarization in degenerate ground states}
\label{sec:polarization_deg}

In the presence of the ground-state degeneracy, we have to be careful
with the Resta-Sorella definition of the polarization.  In particular,
when the ground-state degeneracy is required by the LSM theorem, the
translation operator $T_\alpha$ and the large gauge transformation
$\calU_\alpha$ do not commute (see for example Eq.~\eqref{eq.T_calU.2}).
It immediately follows that, in a translationally invariant ground state
(eigenstate of the translation operator), the expectation
value~\eqref{eq.z} vanishes.  This is due to a kinematical reason, and
does not mean that the system is conductor.  Aligia and
Ortiz~\cite{AligiaOrtiz_PRL1999} proposed to use the expectation value of
$\left( \calU_\alpha \right)^n$, when the ground states are required to
be $n$-fold degenerate by the LSM theorem.  It has a non-vanishing
expectation value in the translationally invariant ground state, from
which one can define the polarization modulo $1/n$.  While their
proposal is interesting and valid, it would lead to a weaker constraint
in our application.  Thus we attempt to utilize the expectation value of
$\calU_\alpha$ itself even in the presence of the ground-state degeneracy.
A physical way to
understand the degeneracy is as follows.  The ground-state degeneracy
can be attributed to the formation of ``charge-density wave'' along the
$\alpha$-direction.  In two (and higher) dimensions, the ground-state
degeneracy can be actually a topological degneracy without any local
order parameter.  However, even in such a case, regarding the system as
a one-dimensional system along the $\alpha$-direction, the ground-state
degeneracy can be attributed to the charge-density wave formation.  This
has been known for example in quantum Hall
states~\cite{TaoThouless_1983,Seidel_PRL2005}.

The translationally invariant ground states are given by
linear superpositions (Fourier transforms) of the
``physical'' charge-density wave ground states.
Each of the physical charge-density wave ground state would have
a definite total polarization, and is related to the other charge-density
wave ground states by the lattice translation $T_\alpha$.
This implies that, each of these states has
a non-vanishing expectation value $\langle \calU_\alpha \rangle$,
which determines the total polarization
via Eqs.~\eqref{eq.z} and~\eqref{eq.P_from_z}.
Furthermore, in each of these states, following
Refs.~\cite{RestaSorella1999,SWM2000,Resta_JPCM2002},
we assume that the fluctuation of the
total polarization $\calP_\alpha$ asymptotically vanishes
in the thermodynamic limit.


\section{Integer Quantum Hall Effect}
\label{sec:IQHE}

\subsection{Quantization of Hall conductance}
\label{sec:IQHE-quantization}

Now let us discuss a simple application of the many-body
polarization and the large gauge invariance.
When a magnetic field perpendicular to the $xy$-plane is applied,
we expect Hall effect, which is characterized by a non-vanishing
Hall conductivity $\sigma_{xy}$.
In a quantum system, under certain conditions,
$\sigma_{xy}$ is quantized to integral (rational)
multiples of $e^2/h$.
This is the celebrated Integer (Fractional) Quantum Hall
Effect.

Laughlin demonstrated~\cite{Laughlin1981}
that the large gauge invariance leads to
the quantization of $\sigma_{xy}$ in IQHE.
This argument was later extended to FQHE~\cite{TaoWu1984}.
However, the original argument involves open boundary conditions,
which accompany gapless edge states.
This complicates theoretical analysis, in particular the
use of the adiabatic process.
An alternative approach based on generalized periodic boundary
condition is later developed by
Niu, Thouless, and Wu~\cite{Niu1985}.
While this approach resolves some of the subtleties of the
Laughlin's argument, the appealing simplicity and physical intuition
of the argument is lost to some extent.
Here, we show that, by considering torus (rectangle with
periodic boundary conditions) without edge states,
Laughlin's gauge invariance argument can be
more directly rephrased in a theoretically better controlled way.

Discussion of IQHE/FQHE is also instructive in introducing a few
operators which will be useful in later discussion.
While we are mostly interested in periodic systems (which are
invariant under discrete translations) in this paper,
we allow more general systems without any translation symmetry,
which may contain impurities (random potentials).
We just assume that the total number of particles is conserved,
the particles couple to the external (electromagnetic)
U(1) gauge field, and
the system has a gap above the ground state.


Now we consider the adiabatic insertion of the magnetic flux
$\Phi$ in the ``hole'' of the torus encircled by the curve in
$2$-direction, over the period of time $T$.
This induces the total ``voltage drop'' in $2$-direction
(integration of the electric field $E_2$)
\begin{equation}
\Delta V_2 = \frac{\Phi}{T}
\end{equation}
during the process.
As a consequence, an electric current
\begin{equation}
 I_1 = \sigma_{xy} \Delta V_2 = \sigma_{xy} \frac{\Phi}{T},
\end{equation}
is induced.
With Eqs.~\eqref{eq.I_calP} and~\eqref{eq.def.calU},
we find
\begin{equation}
 \calF_2(\Phi) \calU_1 = \calU_1 \calF_2(\Phi)
  e^{ 2\pi\imth\frac{\Phi \sigma_{xy}}{e}}.
\label{eq.calF2_calU1}
\end{equation}
In particular, following Laughlin's original argument,
if we choose the inserted flux as $\Phi=\Phi_0$ (unit flux quantum),
\begin{equation}
 \calF_2(\Phi_0) \calU_1 = \calU_1 \calF_2(\Phi_0)
  e^{ 2\pi\imth \frac{\Phi_0 \sigma_{xy}}{e}}
= \calU_1 \calF_2(\Phi_0) e^{2 \pi\imth \tilde{\sigma}_{xy}}.
\end{equation}
On the other hand, the large gauge transformations in two directions
obviously commute:
\begin{equation}
 \calU_1 \calU_2 = \calU_2 \calU_1 .
\end{equation}
It thus follows that
\begin{equation}
 \tilde{\calF}_2(\Phi_0) \calU_1 =
   \calU_1 \tilde{\calF}_2(\Phi_0) e^{2 \pi \imth\tilde{\sigma}_{xy}}.
\label{eq.calF_calU_sigmaxy}
\end{equation}
As we have discussed, $\tilde{\calF}_y(\Phi_0)$ maps a ground state
to a ground state.

Resta and Sorella argued that, insulators can be characterized
by the well-definedness of the electric polarization.
Mathematically, an insulator is characterized by a non-vanishing
ground-state expectation value $\langle \calU_\alpha \rangle$
in the thermodynamic limit, so that the electric polarization
can be defined by its argument (modulo the uncertainty).
Let us assume that the system is in a Quantum
Hall state and thus is an insulator (in the sense that the
diagonal conductivity vanishes).
Furthermore, we assume that the system has a gap
\footnote{
In fact, as is well known, quantization of the Hall conductivity
in a realistic setting requires existence of localized states
due to impurities. These localized states generally give
``gapless'' excitations in terms of energy spectrum, but
they do not respond to flux insertion and do not contribute
to the Hall current.
In such a circumstance, we may extend the notion of ``gapped''
system from a system with a non-vanishing gap to \emph{extended}
states. This subtlety is inherited from Laughlin's original argument
to the present version without edge states.
On the (mathematically) safe side,
we can discuss the Hall conductivity
simply assuming a gap, even though this does not directly address
experimentally observed Quantum Hall Effects.
This is indeed a nontrivial and interesting question, as we can see from
Thouless-Kohmoto-Nightingale-den Nijs paper and its subsequent
generalizations.
}.

Following Resta-Sorella argument, for the unique ground state
$|\Psi_0 \rangle$,
\begin{equation}
 \langle \Psi_0 | \calU_1 | \Psi_0 \rangle = z \neq 0 .
\end{equation}
From the preceding analysis,
$\tilde{\calF}_2(\Phi_0)|\Psi_0 \rangle \sim | \Psi_0 \rangle$
(up to a phase factor), and thus
\begin{equation}
 \langle \Psi_0 |
\left(\tilde{\calF}_2(\Phi_0)\right)^{-1}
\calU_1 \tilde{\calF}_2(\Phi_0) | \Psi_0 \rangle = z \neq 0 .
\end{equation}
Comparing these with Eq.~\eqref{eq.calF_calU_sigmaxy},
we find $\frac{\Phi_0}{e} \sigma_{xy} \in \mathbb{Z}$, namely
the quantization of the Hall conductivity
\begin{equation}
\sigma_{xy} = \frac{e^2}{h} \times \mbox{integer} .
\label{eq.sigmaxy_integer}
\end{equation}
We note that, in this derivation, translation invariance is not
used and therefore the result is applicable to general,
non-periodic systems.

While this is just an alternative formulation of
Laughlin's gauge argument, it is interesting that
the same conclusion can be derived for a system on a torus
without edge states, using Resta-Sorella many-body polarization
operator.
As we will show in the following, the present approach also
has several non-trivial generalizations.


\subsection{IQHE in a periodic system}
\label{sec:IQHE_periodic}

Now let us re-derive the main result~\eqref{thm}
for IQHE in a periodic system, based on the many-body polarization.
All the arguments in Sec.~\ref{sec:IQHE-quantization} still
apply, and thus we have the quantization of the Hall
conductance~\eqref{eq.sigmaxy_integer} holds.
Moreover, we consider the system on a torus of
$L_1 \times L_2$ unit cells,
with $L_1 = q l_1$ and $L_2$ being a coprime with $q$.
The system then has the translation symmetry $T_2$,
and the effective ``translation'' symmetry
$\tilde{\tilde{T}}_1$ as defined in Eq.~\eqref{eq.tiltilT1}.

As in Sec.~\ref{sec:IQHE-quantization}, we assume that the system
is a gapped insulator and has a unique ground state.
Then, the ground state has a non-vanishing expectation value
of $\calU_1$, and the ground state is mapped to itself
by $\tilde{\tilde{T}}_1$.
Thus we immediately find that the phase factor in the commutation
relation~\eqref{eq.tiltilT1_calU1} must be unity.
This implies
\begin{equation}
L_2 \left( \phi \tilde{\sigma}_{xy} - \bar{\rho} \right)
  \in \mathbb{Z}.
\end{equation}
Since this holds for any $L_2$ which is coprime with $q$,
with Eqs.~\eqref{eq.phi_pq}, \eqref{eq.barrho_ppqp},
and~\eqref{eq.rho_q_int}, we find the filling-enforced constraint on the quantized Hall conductivity for IQHE (Theorem \ref{theorem.IQHE}).

\subsection{Fluxon stitching picture}
\label{sec:fluxon_stitching}

We can formulate the flux insertion/large gauge transformation
in a different, more ``local'' manner, by inserting
a flux tube of unit flux quantum $\Phi_0$
orthogonal to the two-dimensional system, and then
dragging it across the system~\cite{Wen1990b}.
After ``stitching'' the system by dragging the flux tube along
a fundamental cycle of the torus, the ``hole'' of the torus
corresponding to the other fundamental cycle acquires the
unit flux quantum $\Phi_0$.
Thus we can expect that this process is topologically equivalent
to the adiabatic insertion of the unit flux quantum
$\calF_\alpha$ introduced earlier.

More precisely, this process can be defined as follows.
First we define the operator $f_\alpha(\vec{r})$
which corresponds to an ``adiabatic'' increase of
vector potential in $\alpha$ direction by $\Phi_0/l_\alpha$,
over the local region of $l_1 \times l_2$ unit cells
located at $\vec{r}$.
When we apply $f_\alpha(\vec{r})$ to a ground state
(vacuum) $|\Psi_0\rangle$,
$f_\alpha(\vec{r})|\Psi_0\rangle$, a flux tube with
the unit flux quantum $\Phi_0$ piercing through two-dimensional
system is created on one side of $\vec{r}$,
and another one with $-\Phi_0$ is created on
the side of $\vec{r}$.
The adiabatic increase of the vector potential induces
an elecric field, which generates Hall current.
As a result, the charge $\pm \tilde{\sigma}_{xy}$ is induced
where the flux tubes go through the system.
We call the composite object of the charge and the flux tube
as a ``fluxon''.
Thus $f_\alpha(\vec{r})$ can be regarded as a creation
operator of the fluxon-antifluxon pair.
In fact, because $f_\alpha(\vec{r})$ creates an excited state
with a fluxon-antifluxion pair, this process is actually non-adiabatic.
However, we assume that, by increasing the vector potential
sufficiently slowly, it does not create any excitation other than
those required topologically, and is ``adiabatic'' in a generalized
sense.
Now, by further applying
$f_{\alpha}(\vec{r}- l_{\bar{\alpha}}\vec{e}_{\bar{\alpha}})$,
we can move the fluxon by $-l_{\bar{\alpha}} \vec{e}_{\bar{\alpha}}$.
Thus, the consecutive product
\begin{equation}
\frakF_\alpha(\vec{r})
\equiv \prod_{j=L_{\bar{\alpha}}/l_{\bar{\alpha}}}^0
     f_\alpha(\vec{r} - j l_{\bar{\alpha}}) ,
\end{equation}
when acted on the ground state $|\Psi_0 \rangle$,
creates a fluxon-antifluxon pair, drags the fluxon along the
fundamental cycle of the torus in $\bar{\alpha}$-direction,
and pair-annihilate the fluxon and antifluxon.
Thus the resulting state should be a ground state (vacuum),
as long as the process is ``adiabatic'' in the extended sense
discussed above.
However, it should be also noted that
$\frakF_\alpha(\vec{r})$ leaves a string of
modified vector potential behind the path, and thus
the Hamiltonian after the process is not identical to the initial one.
As in the case of $\calF_\alpha$, this modified vector potential
can be eliminated by a large gauge transformation, but
it has to be squeezed onto the affected region.
Namely, we define
\begin{equation}
 \frakU_\alpha(\vec{r}) =
 \exp{\left( i
 \int d\vec{r}' \; \theta_{\vec{r}}(\vec{r}') \rho(\vec{r}') \right)},
\end{equation}
where
\begin{equation}
  \theta(\vec{r}')  =
     \begin{cases}
      0 & r'_{\alpha} <  r_{\alpha}  \\
     2 \pi \frac{r'_{\alpha}- r_\alpha}{l_\alpha} &
      r_{\alpha} \leq r'_{\alpha} <  r_{\alpha}+ l_{\alpha}  \\
      2 \pi & r_{\alpha} + l_{\alpha} < r'_{\alpha}
     \end{cases} .
\end{equation}
Then
\begin{equation}
 \tilde{\frakF}_{\alpha}(\vec{r}) \equiv
  \frakU_{\alpha}(\vec{r}) \frakF_{\alpha}(\vec{r})
  \end{equation}
maps a ground state of the original Hamiltonian to a
ground state of the same Hamiltonian.

Similarly to Eq.~\eqref{eq.T_calU},
\begin{equation}
 T_\alpha \frakU_{\alpha}(\vec{r}) {T_\alpha}^{-1}
{\frakU_{\alpha}(\vec{r})}^{-1}
=e^{-\frac{2\pi\imth}{l_\alpha}
\int_{r_\alpha < r'_\alpha \leq r_\alpha + l_\alpha}
d\vec{r}' \; \rho{(\vec{r}')}} .
\end{equation}
The integral in the exponent gives the total particle number
within the strip
$r_\alpha < r'_\alpha \leq r_\alpha + l_\alpha$.
It is not a conserved quantity like the total particle
number in the entire system.
However, assuming an incompressible liquid state,
the particle number within the strip has only small
fluctuations if $L_{\bar{\alpha}}$ is sufficiently larger
than the correlation length of the pair correlation function
(density-density correlation function).
Thus it may be replaced as
\begin{equation}
\int_{r_\alpha < r'_\alpha \leq r_\alpha + l_\alpha}
d\vec{r}' \; \rho{(\vec{r}')}  \sim \bar{\rho} l_\alpha L_{\bar{\alpha}} .
\end{equation}
Within this assumption,
\begin{equation}
 T_\alpha \frakU_{\alpha}(\vec{r}) {T_\alpha}^{-1}
{\frakU_{\alpha}(\vec{r})}^{-1} \sim
e^{-{2\pi\imth\bar{\rho} L_{\bar{\alpha}}}}.
\end{equation}
Thus we expect
\begin{equation}
 {T_\alpha} \tilde{\frakF}_{\alpha}(\vec{r}) {T_\alpha}^{-1}
 \sim \tilde{\frakF}_{\alpha}(\vec{r})
e^{-{2\pi\imth\bar{\rho} L_{\bar{\alpha}}}},
\label{eq.T_frakF}
\end{equation}
when acting on the ground-state subspace.

Let us consider the system on a torus consisting of
$L_1 \times L_2$ unit cells,
with the Landau gauge~\eqref{eq.LandauGauge}.
As in Sec.~\ref{sec:eff_sym},
we take $L_1$ to be an integral multiple of $q$, and
$L_2$ to be a coprime with $q$.
Here we consider the ``fluxon stitching'' in  $2$-direction,
$\frakF_1(\vec{r}_0)$, with the starting point $\vec{r}_0$.
In the present setup with the Landau gauge~\eqref{eq.LandauGauge},
the system is not invariant under $T_1$, and
it is even impossible to define the magnetic translation operator
in $1$-direction.
Nevertheless, as we have discussed in Sec.~\ref{sec:IQHE_periodic},
we can define an effective symmetry generator~\eqref{eq.tiltilT1}
on the ground-state subspace.
Similar to Eq.~\eqref{eq.calF2_calU1},
\begin{equation}
 \calF_2(\varphi \Phi_0) \frakU_1(\vec{r}) =
\frakU_1(\vec{r}) \calF_2(\varphi \Phi_0 )
  e^{ 2\pi i \varphi \tilde{\sigma}_{xy}} .
\label{eq.calF2_frakU1}
\end{equation}
Combining Eq.~\eqref{eq.T_frakF} and
Eq.~\eqref{eq.calF2_frakU1} with $\varphi = \phi L_2$,
we find
\begin{equation}
\big(\tilde{\tilde{T}}_1\big)^{-1}\big[\tilde{\frakF}_1(\vec{r})\big]^{-1} \tilde{\tilde{T}}_1  \tilde{\frakF}_1(\vec{r}) \sim
  e^{2 \pi i L_2 (\phi \tilde{\sigma}_{xy} - \bar{\rho} )}\sim1
\label{eq.tiltilT1_frakF1}
\end{equation}
Now, both $\tilde{\frakF}_1$ and $\tilde{\tilde{T}}_1$
maps the ground state to the ground state.
Because of the uniqueness of the ground state (which was assumed),
the phase factor appearing above should be unity.
Thus we recover Theorem 1, using the fluxon stitching picture.
We will see that this formulation also provides a useful insight
for Fractional Quantum Hall effect.

\section{Fractional Quantum Hall effects}
\label{sec:FQHE}

\subsection{Fractional quantization of the Hall conductivity}
\label{sec:FQH_quantization}

A FQHE must accompany a topological ground-state
degeneracy on a torus~\cite{TaoWu1984}.
As we have discussed in Section~\ref{sec:polarization_deg},
we can choose ``charge-density wave'' ground state
with respect to the direction under consideration,
so that the total polarization of the ground in that
direction is well-defined (expectation value of $\calU_1$
does not vanish).
In fact, in our setup with the Landau gauge~\eqref{eq.LandauGauge},
the system does not have the one-site translation symmetry $T_1$
in $1$-direction. Thus we can expect that a charge-density wave
ground state in $1$-direction is automatically chosen as the
ground state of a generic finite-size system.

The commutation relation~\eqref{eq.calF_calU_sigmaxy} then
implies that, application of $\tilde{F}_2(\Phi_0)$ changes
the expectation value of $\calU_1$ by the phase factor
$e^{ 2\pi\imth\tilde{\sigma}_{xy}}$.
In other words, the total polarization $\bar{\calP}_1$
is changed by $\tilde{\sigma}_{xy}$.
This implies that, if $\tilde{\sigma}_{xy}$ is fractional and
the system has a gap,
then the new ground state obtained by the adiabatic
process $\tilde{F}_2(\Phi_0)$ is different from the initial
ground state.
Thus, if the Hall conductivity takes a fractional value
and the system has a gap above the ground states,
\begin{equation}
 \tilde{\sigma}_{xy} = \frac{\tilde{p}}{\tilde{q}}
\end{equation}
for coprimes $\tilde{p}$ and $\tilde{q}$, there must be
$\tilde{q}$ degenerate, independent ground states.
Namely, if there are (only) $n$ degenerate
ground states below the gap, the Hall conductivity is
quantized in unit of $(1/n)(e^2/h)$.
In other words, the ground state degeneracy $n$ must
be an integral multiple of $\tilde{q}$.

This is the ``non-edge'' version of the argument by
Tao and Wu~\cite{TaoWu1984}.
The present argument has the advantage particularly
in the fact that the ground-state degeneracy is better
defined in the absence of the gapless edge states.

\subsection{FQHE in a periodic system}
\label{sec:FQHE_periodic}

Now let us consider a many-particle system in a periodic
potential or on a periodic lattice, with the total
particle number conserved.
As we have discussed in Section~\ref{sec:mom_count_torus},
under the magnetic field,
the particle number per the \emph{magnetic} unit cell,
$\bar{\rho} q$ is the relevant parameter for the LSM theorem.
That is, if
\begin{equation}
q \bar{\rho} = \frac{p''}{q''},
\label{eq.def_qpp}
\end{equation}
is fractional with mutually coprime
$p''$ and $q''$, the system must be either gapless
or has $q''$-fold ground-state degeneracy.

Similarly to the case of the IQHE, we can obtain a stronger constraint
based on the many-body polarization.
As in Sec.~\ref{sec:IQHE_periodic}, we consider the system on the
torus of the size $L_1 \times L_2$ with the
Landau gauge~\eqref{eq.LandauGauge}, and choose
$L_1$ to be an integral multiple of $q q'$ and
$L_2$ to be a coprime with $q$ and $q'$.
Then Eq.~\eqref{eq.tiltilT1_calU1} still holds.
As in Sec.~\ref{sec:FQH_quantization}, we assume that
the system is an insulator in Resta-Sorella sense.
That is,
we can choose a complete set of the degenerate ground states,
so that each of which has a non-vanishing expectation value
$\langle \calU_1 \rangle$ with an asymptotically vanishing
fluctuation of its argument, in the thermodynamic limit.
Starting one particular ground state with a well-efined
expectation value $\langle \calU_1 \rangle$,
Eq.~\eqref{eq.tiltilT1_calU1} implies that
we can generate another ground state which
has the total polarization $\calP_1$ different from
the original ground state by
\begin{equation}
 2 \pi  L_2 \left( \phi \tilde{\sigma}_{xy} - \bar{\rho} \right) .
\end{equation}
Since the total polarization is defined modulo $2 \pi$,
and and we can choose  $L_2$ as any integer,
we can generate $\frakq$ independent ground states
(including the original one) if
\begin{equation}
\phi \tilde{\sigma}_{xy} - \bar{\rho}  = \frac{\frakp}{\frakq} .
\end{equation}
This implies that the number of the degenerate ground states,
$n$, must be an integral multiple of $\frakq$.
Thus
\begin{equation}
  n \left( \phi \tilde{\sigma}_{xy} - \bar{\rho} \right) \in
    \mathbb{Z} ,
\label{eq.FQHE_GSdeg_thm}
\end{equation}
or equivalently, using Eq.~\eqref{eq.def_qpp}
\begin{equation}
p \frac{\tilde{p}}{\tilde{q}} - \frac{p''}{q''}
\in \frac{q}{n} \mathbb{Z} .
\end{equation}
As a corollary,
this relation includes the fact that the ground-state degeneracy
$n$ is an integral multiple of both $\tilde{q}$ and $q''$.
The fact that $n$ is an integral multiple of $q''$
follows from the generalized LSM, and that $n$ is an integral
multiple of $\tilde{q}$ follows from Ref.~\cite{TaoWu1984}.
However, the present result puts a stronger constraint
in terms of the relation~\eqref{eq.FQHE_GSdeg_thm}.





\subsection{Fluxon stitching and braiding with background anyons}
\label{sec:fluxon braiding}

Let us now discuss the fluxon stitching picture introduced
in Sec.~\ref{sec:fluxon_stitching}, for
the Fractional Quantum Hall states.
All the discussions leading to Eq.~\eqref{eq.tiltilT1_frakF1}
remain valid.
The difference arises because the ground state is no longer
unique, the phase factor appearing in Eq.~\eqref{eq.tiltilT1_frakF1}
is not necessarily unity.
In fact, it is generally not unity, and the nontrivial phase
factor is related to the ground-state degeneracy, as
we have just discussed in Sec.~\ref{sec:FQHE_periodic}.
On the other hand, the fluxon stitching picture provides
an intuitive picture on the physical origin of the nontrivial
phase factor.

\begin{figure}
\includegraphics[width=0.9\columnwidth]{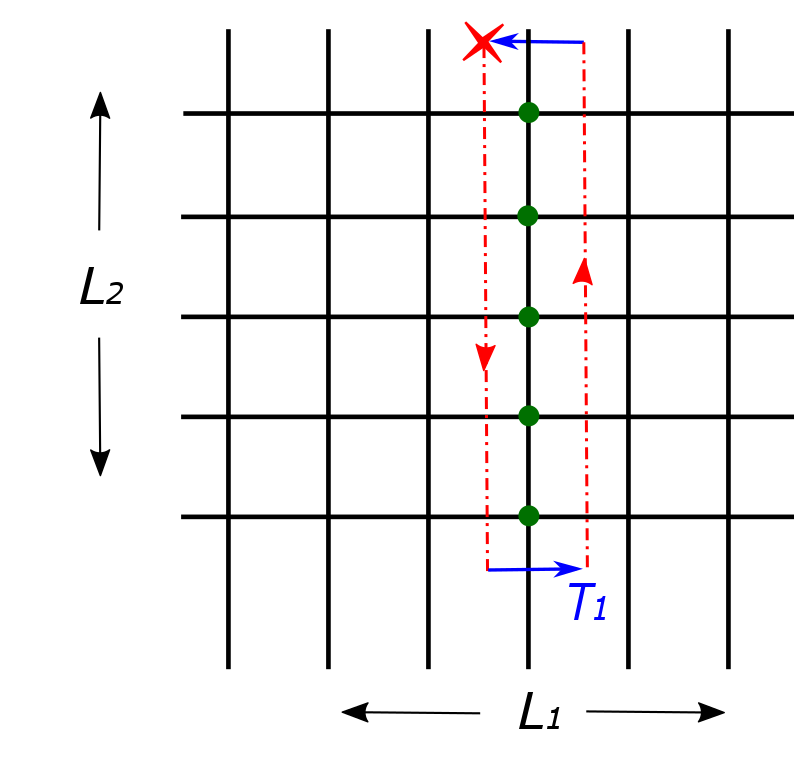}
\caption{(color online) An illustration of the fluxon excitation (denoted by red cross) and the fluxon braiding process corresponding to
Eq.~\protect\eqref{eq.tiltilT1_frakF1_phase}.}
\label{fig:vison string}
\end{figure}

That is, Eq.~\eqref{eq.tiltilT1_frakF1} can be rewritten as
\begin{equation}
e^{2\pi\imth\bar\rho L_2}\sim\big(T_1\big)^{-1} \tilde{\frakF}_1 T_1 \big(\tilde{\frakF}_1\big)^{-1}
\sim \mathcal{F}_2\tilde{\frakF}_1 \big(\mathcal{F}_2\big)^{-1}\big(\tilde{\frakF}_1\big)^{-1}
\label{eq.tiltilT1_frakF1_phase}
\end{equation}
where we have used short-handed notation $\mathcal{F}_2\equiv\mathcal{F}_2(\phi L_2\Phi_0)$ and $\tilde{\frakF}_1\equiv \tilde{\frakF}_1({\vec r})$. The left-hand side of this equation can be interpreted as
the probability amplitude of the process of moving the fluxon counter-clockwise
along the closed loop as shown in Fig.~\ref{fig:vison string}. This closed loop encircles exactly one column of the torus in FIG. \ref{fig:vison string}. The adiabatic Berry phase acquired in this ``fluxon braiding'' process is $e^{2\pi\imth\bar\rho L_2}$ as directly computed in (\ref{eq.T_frakF}), and it holds true irrespective of integer or fractional QHEs. On the other hand, the right hand side of (\ref{eq.tiltilT1_frakF1_phase}) can be decomposed into two contributions:
\bea\notag
&\mathcal{F}_2\tilde{\frakF}_1 \big(\mathcal{F}_2\big)^{-1}\big(\tilde{\frakF}_1\big)^{-1}\sim\\
&\big[\mathcal{F}_2{\frakU}_1 \big(\mathcal{F}_2\big)^{-1}\big({\frakU}_1\big)^{-1}\big]\cdot
\big[\mathcal{F}_2{\frakF}_1 \big(\mathcal{F}_2\big)^{-1}\big({\frakF}_1\big)^{-1}\big]\label{two terms on rhs}
\eea
The 1st contribution
\bea\label{AB phase}
\mathcal{F}_2{\frakU}_1 \big(\mathcal{F}_2\big)^{-1}\big({\frakU}_1\big)^{-1}\sim e^{2\pi\imth\tilde\sigma_{xy}\phi L_2}
\eea
can be understood as an Aharonov-Bohm phase, acquired when a fluxon carrying charge $\tilde\sigma_{xy}$ goes around $\phi L_2$ flux quanta piercing through each column of the torus in FIG. \ref{fig:vison string}. This is true for both integer and fractional QHEs.

Meanwhile, the 2nd term $\mathcal{F}_2{\frakF}_1 \big(\mathcal{F}_2\big)^{-1}\big({\frakF}_1\big)^{-1}$ corresponds to nothing but the mutual braiding statistics between two particles\cite{Wen1990b,Oshikawa2006}: one is a fluxon $F$ and the other is the ``background anyon''\cite{Cheng2016} contained in each column, as we will elaborate soon. In the case of Integer Quantum Hall states, no anyon exists and this braiding phase is unity, and therefore we recover \eqref{eq.tiltilT1_frakF1} and Theorem \ref{theorem.IQHE}. The nontrivial phase appearing in the case of a Fractional Quantum Hall state can be understood as follows.

Quite generally in a topologically ordered system on a lattice, in the presence of translational symmetry, the
spatial arrangement of anyons in the ground state forms a translational-invariant periodic structure, known as fractionalization of translation symmetries (see Section \ref{sec:frac_mag_trans} for details)~\cite{Essin2013,Barkeshli2014,Tarantino2016,Cheng2016}. In particular one (Abelian) anyon $a$ can be placed in each unit cell without breaking lattice translational symmetry, which we refer to as the ``background anyon'' $a$. When another anyon $b$ travels around each unit cell counter-clockwise for once, it will acquire a braiding phase $e^{\imth\theta_{b,a}}$ in this process. Since the fluxon $F$ goes around the array of the background anyons in a full column,
this process picks up the phase factor
\begin{eqnarray}
 \mathcal{F}_2{\frakF}_1 \big(\mathcal{F}_2\big)^{-1}\big({\frakF}_1\big)^{-1}\sim e^{\imth\theta_{F,a} L_2},
\label{phase:anyon}
\end{eqnarray}
due to the fractional statistics between the fluxon and the background anyons.
Here $\theta_{F,a}\in[0,2\pi)$ represents the mutual statistical angle
between fluxon ($F$) and background anyon ($a$) in each unit cell. Note
that although distinct anyons can be permuted by translation
symmetry\cite{Kitaev2006,Wen2003,Barkeshli2012b,Cheng2016}, their mutual
statistics with fluxon $F$ should be invariant under translation so
$\theta_{F,a}$ is well-defined.

Comparing Eqs.~\eqref{eq.tiltilT1_frakF1_phase}, (\ref{two terms on rhs}) and~\eqref{phase:anyon},
we find Theorem~\ref{theorem.FQHE}.
Furthermore, comparing this with Eq.~\eqref{eq.FQHE_GSdeg_thm},
we find the relation~\eqref{eq.thetaFa_gsd}
between the statistical angle and the ground-state degeneracy.

In Appendix \ref{app:fluxon braiding} we explicitly computed the Berry phase of the fluxon moving process in \eqref{eq.tiltilT1_frakF1_phase} in the effective Abelian Chern-Simons field theory, therefore establishing \eqref{eq.tiltilT1_frakF1_phase}-\eqref{phase:anyon} in all Abelian topological orders. While such a field theory calculation does not apply to non-Abelian FQHEs, an alternative argument based on symmetry fractionalization algebra proves Theorem \ref{theorem.FQHE} for both Abelian and non-Abelian FQHEs, as we will show in the following section.

\section{Fractionalization algebra of magnetic translation symmetry}
\label{sec:frac_mag_trans}

In this Section, we discuss the problem from the viewpoint of
symmetry fractionalizations~\cite{Essin2013,Barkeshli2014,Tarantino2016},
which gives a modern understanding of topological order.
In particular, this is useful in describing systems with
a non-Abelian topological order.

\begin{figure}
\includegraphics[width=0.9\columnwidth]{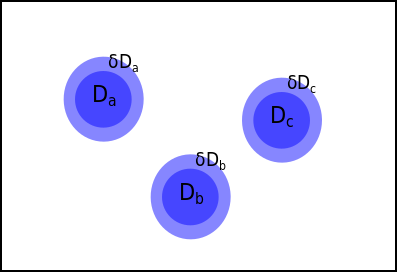}
\caption{(color online) The fractionalization of global symmetries into spatially local regions.}
\label{fig:fractionalization}
\end{figure}

Let us consider a general gapped quantum phase (which may be topologically ordered) respecting a global symmetry group $SG$, and assume that the symmetry transformation would \emph{not} exchange anyon types (\ie superselection sectors) in the system. If one wants to understand how a global symmetry element $g\in SG$ transformes local excitations within finite spatial regions $D_a,D_b$ far apart from each other, one could decompose global symmetry transformation $R_g$ on the full many-body Hilbert space into a direct product of local symmetry actions $\Omega_g(a),\Omega_g(b)...$. For example, $\Omega_g(a)$ only acts on a region $D_a\cup \delta D_a$, which is slightly larger than $D_a$ by a ribbon $\delta D_a$ (with a width $\sim$ correlation length) around the edge of $D_a$ (see Fig.\ref{fig:fractionalization}). These local operators $\Omega_g(a),\Omega_g(b)...$ will be called as fractionalized symmetry operators below.

The basic assumption of symmetry fractionalization is the following condition~\cite{Essin2013,Barkeshli2014,Tarantino2016}:
\begin{align}
 R_g|\psi(D_a,D_b...)\rangle=e^{i\theta_g}\cdot \Omega_g(a)\cdot\Omega_g(b)...|\psi(D_a,D_b...)\rangle,\label{eq:fractionalization}
\end{align}
where $|\psi(D_a,D_b...)\rangle$ describes any states whose quasiparticles are only localized within regions $D_a\cup D_b..$, including the ground state. Here $e^{i\theta_g}$ is a possible $U(1)$ phase factor due to the background $g$-quantum-number outside the regions $D_a\cup D_b..$. Importantly, $\theta_g$ is independent of the excitations inside regions $D_a\cup D_b..$.

Next, let us focus on one particular region, e.g. $D_a$. To save notation we denote this region as $D$ (and similarly denote the ribbon area around it as $\delta D$). Physically, for onsite unitary symmetry $g$, such as the $U(1)$ charge conservation, $\Omega_g$ on $D\cup \delta D$ can be obtained by the following adiabatic process. $g$ being onsite means that $R_g=\prod U_g(\vec r)$ is a direct product of onsite symmetry operations. First, one creates a pair of $g$-symmetry defects in $\delta D$ and adiabatically braid them around $D$ along $\delta D$ then annihilate them (i.e., $\delta D$ can be viewed as the worldline of the $g$-defect). Second, one applies $\prod_{\vec r\in D} U_g(\vec r)$ within the region $D$. The combination of these two operations well-defines the transformation of any excitated states in $D$, and sends the ground state back to the ground state (up to a $U(1)$ phase factor). One may use this combination as a construction of $\Omega_g$ for onsite symmetry $g$.

Note that by adiabatically creating $g$-symmetry defects, we mean adiabatically modifying the Hamiltonian such that there exist a string in the real space, like a branch cut. Any term in the Hamiltonian crossing the string is modified such that only the degrees of freedom on \emph{one side} of the string is conjugated by $R_g$. The end points of the string are the $g$-symmetry defects. After braiding the $g$-symmetry-defect around $\delta D$, the final Hamiltonian is different from the original Hamiltonian by $\prod_{\vec r\in D} U^{-1}_g(\vec r)$ conjugation, and applying $\prod_{\vec r\in D} U_g(\vec r)$ would send this final Hamiltonian back to the original Hamiltonian.

For time-reveral and spatial symmetries, such as the (magnetic) translation symmetry studied here, similar defect-based construction of $\Omega_g$ becomes difficult. For instance, the symmetry-defect associated with translational symmetry is the dislocation, braiding which around a region involves adding/removing physical degrees of freedom. However, conceptually one still can accept the existence of $\Omega_g$ satisfying \ref{eq:fractionalization} quite generally (as long as $g$ is not changing the anyon types or superselection sectors). In fact, $\Omega_g$ for (magnetic) translational symmetries can be microscopically constructed if one employs certain formulations such as tensor-network or parton methods~\cite{Jiang2017}.

Below we assume that $\forall g$ in the magnetic translation symmetry group, $\Omega_g$ exists and they satisfy Eq.(\ref{eq:fractionalization}). Note that in order to satisfy Eq.(\ref{eq:fractionalization}), $\Omega_g$ is not completely fixed. We already mentioned that $\Omega_g$ may include an overal $U(1)$ phase when acting on all excited states. But what is more important is that $\Omega_g$ may include a nontrivial phase factor dependent on the nature of the excitation: because the only modifiable action of $\Omega_g$ is within $\delta D$, which is spatially separated with the excitations, this dictates that the only nontrivial ambiguity is a braiding of an Abelian anyon $\epsilon(g)$ around $\delta D$~\cite{Barkeshli2014,Tarantino2016}.

Namely, combining $\Omega_g$ with an additional $\epsilon(g)$-anyon braiding: $\Omega_g\rightarrow \Omega_g'=\epsilon(g)\cdot\Omega(g)$ can give another well-defined $\Omega_g'$ action, which still satisfy Eq.(\ref{eq:fractionalization}) (after including the effect of $\epsilon(g)$ in other regions). Interestingly, when nontrivial symmetry fractionalization occurs, it is impossible to avoid the anyon braiding in the ambiguity of $\Omega_g$. Precisely, the symmetry fractionalization pattern in a gapped quantum phase is determined by the following algebra:
\begin{align}
 \forall g_1,g_2\in SG,\;\Omega_{g_1}\cdot\Omega_{g_2}=\lambda(g_1,g_2) \Omega_{g_1\cdot g_2},\label{eq:frac_lambda}
\end{align}
where $\lambda(g_1,g_2)\in \mathcal{A}$ is an Abelian anyon and the Abelian group $\mathcal{A}$ is the fusion group of all Abelian anyons in the system. Associativity dictates that:
\begin{align}
 \lambda(g_1,g_2)\cdot\lambda(g_1\cdot g_2,g_3)=\lambda(g_2,g_3)\cdot\lambda(g_1, g_2\cdot g_3).\label{eq:frac_2_cocycle}
\end{align}
Together with the $\epsilon(g_1),\epsilon(g_2),\epsilon(g_3)$ ambiguities, this indicates $\lambda(g_1,g_2)\in H^{2}(SG,\mathcal{A})$, i.e., $\lambda(g_1,g_2)$ is an element in the second cohomology group of $SG$ with coefficient in $\mathcal{A}$ (with trivial group action because we are considering the case with anyon types invariant under $SG$).

Let us now apply this general framework to the magnetic translation group in the presence of global $U(1)$ charge conservation. The fundamental Eq.(\ref{mag trans}) indicates that:
\begin{align}
\Omega_{\tilde T_1}\Omega_{\tilde T_2}\Omega^{-1}_{\tilde T_1}\Omega^{-1}_{\tilde T_2}=\lambda_a \Omega_{U(\phi)},\label{eq:frac_mag_transl}
\end{align}
where $U(\phi)\equiv e^{2\pi i \phi \hat N}$ is the global $U(1)$ rotation by angle $2\pi \phi$. And $\lambda_a$ is the braiding of certain anyon $a$ around $\delta D$. This anyon-$a$ is appearing in the Theorem \ref{theorem.FQHE}. Note that $\Omega_{\tilde T_1}$ and $\Omega_{\tilde T_2}$ have their overall $U(1)$ phases, $\epsilon(\tilde T_1)$ and $\epsilon(\tilde T_2)$ ambiguities, but the product on the LHS of Eq.(\ref{eq:frac_mag_transl}) has all these ambiguities cancelled. This product and consequently the RHS, should be a completely fixed action on the ground state and excitations in the region $D$.

Although the RHS of Eq.(\ref{eq:frac_mag_transl}) is completely determined, the anyon-$a$ and $\Omega_{U(\phi)}$ could have a relative ambiguity. Let us recall one definition of $\Omega_{U(\phi)}$, based on $U(\phi)$-defect construction. Because $U(1)$ group is continuous, this adiabatic construction can be made in a long length scale and low energy scale so that any irrelevant anyons will not be excited. Namely, the adiabatic insertion of a $\phi$-flux over a large area (but within $\delta D$), which is nothing but the $U(\phi)$-defect, should give a well-defined state.  If we use this this natural definition of $\Omega_{U(\phi)}$, then $a$ is determined as long as $\phi$ is determined.

The subtlety is that in the global magnetic translation algebra Eq.(\ref{mag trans}), only the fractional part of $\phi$ is well-defined. Therefore, it is perfectly fine to use $\phi'=\phi+1$ in the global magnetic translation. However, in the fractionalized operator $\Omega_{U(\phi)}$, adiabatically inserting a $\phi$-flux and $\phi'$-flux over a large area would generally result in different defects. Clearly, $U(\phi')=U(\phi)+\lambda_F$ where $F$ is a particular anyon in the system which is created by adiabatically inserting $2\pi$ flux over a large area. And $\lambda_F$ is the braiding operation of $F$ around $\delta D$. Consequently, because the RHS of Eq.(\ref{eq:frac_mag_transl}) is fixed, if $\phi\rightarrow \phi+1$, then we must have $\lambda_a\rightarrow \lambda_{a-F}$ where $a-F$ is the anyon obtained by fusing $a$ and the antiparticle of $F$. This ambiguity has been discussed below Theorem-2.

Theorem \ref{theorem.FQHE} turns out to be just an application of Eq.(\ref{eq:frac_mag_transl}). Let us consider an anyon $F$ sitting inside region $D$. The LHS and RHS of Eq.(\ref{eq:frac_mag_transl}) describe two different actions on this state. Relative to the ground state, the RHS would pick up two phase factors: the braiding phase factor $\theta_{F,a}$ and the phase factor obtained by $\Omega_{U(\phi)}$. But $\Omega_{U(\phi)}$ is nothing but a measurement of the charge carried by $F$ by $\phi$-$U(1)$ rotation. Because $F$ must carry charge $\tilde\sigma_{xy}$, we know that this second phase factor is simply $e^{2\pi\imth\phi\tilde\sigma_{xy}}$.

Now let us look at the LHS, which magnetic translate the unit-fluxon-$F$ around one unit cell. The phase obtained in this process can be safely shown to be due to the charge density per unit cell: $e^{2\pi\imth\bar \rho}$. The details of the derivation can be found in Appendix~\ref{app:mag_trans_fluxon}. Summarizing our results, we have obtained Theorem~\ref{theorem.FQHE}
\begin{align}\label{fqh thm}
 e^{2\pi\imth\bar \rho}=\mbox{LHS}=\mbox{RHS}=e^{2\pi\imth\phi\tilde\sigma_{xy}+\imth\theta_{F,a}}.
\end{align}

\section{Generalization to non-symmorphic lattices}
\label{sec:non-symmorphic}

Now let us generalize theorem (\ref{thm}) to a 2d non-symmorphic lattice with glide reflections. Among all 17 space groups in 2d (wallpaper groups) only 4 are non-symmorphic\cite{Dresselhaus2008B}: $pg$ and $pmg$ with one glide plane, $pgg$ and $p4g$ with two perpendicular glide planes. In a 2d system $\sigma_{xy}$ is odd under orientation-reversing glide operation, so we consider anti-unitary ``magnetic glide'' operation $\tilde g=\calU\cdot g\cdot\bst$ which includes a time reversal operation $\bst$, where $\calU$ is certain unitary transformation responsible for the magnetic translation algebra (\ref{mag trans}). Specifically we focus on the following magnetic algebra between magnetic glide $\tilde g_1$ and magnetic translation $\tilde T_2$
\bea\label{mag glide algebra}
\tilde T_2~\tilde g_1~\tilde T_2~(\tilde g_1)^{-1}=e^{-\imth\pi\phi\hat N}
\eea
where $\hat N=\sum_{\bf r}\hat N_{\bf r}$ is the total charge of the system. Choosing a Landau gauge where $\tilde T_2=T_2$ is the usual translation along $\vec a_2$ direction, we have
\begin{eqnarray}\label{glide}
&\tilde g_1\equiv e^{\imth\pi\phi\sum_{\bf r}r_2\hat N_{\bf r}}\cdot g_1\cdot\bst,\\
&\notag(r_1,r_2)\overset{g_1}\longrightarrow(r_1+1/2,-r_2),~~~(\tilde g_1)^2=\tilde T_1.
\end{eqnarray}
where ${\bf r}=r_1\vec a_1+r_2\vec a_2$ again denotes the center of mass position for each unit cell. It's a convenient representation of magnetic translation (\ref{mag trans}) on a 2d non-symmorphic lattice, with $\phi$ flux quanta in each unit cell.

Completely in parallel to theorems (\ref{thm}) and (\ref{thm:fractional}), we can obtain the following relation based on fractionalization algebra of (\ref{mag glide algebra}) acting on a single fluxon $F$:
\bea\label{fqh glide}
e^{\imth\frac{\bar\rho}2}=e^{\imth\tilde\sigma_{xy}\frac{\phi}2}\cdot e^{\imth\theta_{F,a}}
\eea
The difference between the above relation and (\ref{fqh thm}) of the magnetic translation case can be intuitively understood as the following. Magnetic glide operation $\tilde g_1$ can be viewed as a half of a magnetic translation $\tilde T_1$, dividing each unit cell into two halves. With glide symmetry, the charge density $\bar\rho/2$ and flux density $\phi\Phi_0/2$ in each half of a unit cell becomes well defined, and so is the background anyon $a$ in each half of the unit cell. According to (\ref{fqh glide}), we immediately prove the following theorem
\begin{theorem} {\bf Filling-enforced constraint for QHE on non-symmorphic lattices:}
Hall conductivity of a gapped 2d insulator must satisfy
\begin{eqnarray}
\bar\rho=\sigma_{xy}\cdot\phi+\frac{\theta_{F,a}}{\pi}\mod2.
\label{thm:nonsymmorphic:fractional}
\end{eqnarray}
if it preserves magnetic algebra (\ref{mag glide algebra}) on a non-symmorphic lattice preserving glide symmetry (\ref{glide}).
\label{theorem.FQHE.ns}
\end{theorem}
An immediate corollary for $\phi=0$ is that for a generic system preserving magnetic glide symmetry (\ref{glide}), \ie the combination of glide $g_1$ and time reversal $\bst$ (where arguments in \Ref{Parameswaran2013} does not apply), charge density $\bar\rho$ per u.c. must be an even integer to achieve a non-fractionalized featureless insulator.

\section{Applications}\label{sec:applications}

\subsection{IQHE models}

Theorems \ref{theorem.IQHE} and \ref{theorem.FQHE} directly apply to
integer and fractional QHEs on lattices, \ie integer and fractional
Chern insulators.
We first discuss applications of
Theorem \ref{theorem.IQHE} to IQHEs with a unique ground state
on a torus.

The Harper-Hofstadter models~\cite{Harper1955,Hofstadter1976} which
study the motion of Bloch electrons in a magnetic field is a simple
realization of nontrivial IQHE~\cite{TKNN}.  While
Theorem~\ref{theorem.IQHE} was proved earlier in Ref.~\cite{Dana1985},
apparently its consequences has not been much appreciated.  $\pi$-flux
models ($\phi=\frac12$) of spinless particles at half filling (half
charge per lattice site) has been studied in various cases. Theorem
\ref{theorem.IQHE} dictates that with $\bar\rho=\frac12$ per unit cell,
\eg on square, triangular and kagome lattices, any unique ground state
must have an \emph{odd} Hall conductance $\tilde\sigma_{xy}=1\mod2$. In
particular, $\tilde{\sigma}_{xy}$ cannot vanish in this case, even
though the time-reversal symmetry of $\pi$-flux models seems to allow
$\tilde{\sigma}_{xy}=0$.  Of course it is still possible to have
$\tilde{\sigma}_{xy}=0$ in the $\pi$-flux models. However, for that
there must be degenerate ground states.  In other words, a trivial
insulator at half filling is forbidden.
Now with our proof in the present paper, this statement also applies to
interacting systems.

In honeycomb-lattice $\pi$-flux models at half filling,
we have $\bar\rho=1$.
Then Theorem~\ref{theorem.IQHE} requires an \emph{even}
Hall conductance, provided that the ground state is unique.
This is indeed the case for IQH states of free fermions~\cite{Bercx2014}.
In fact, our proof is valid also for interacting bosons, and
requires an even Hall conductance for a bosonic model
on the honeycomb lattice with $\pi$-flux.
This is indeed consistent with the finding of a recent work~\cite{He2015a}.

\subsection{Magnetically ordered systems}

Another less obvious application is magnetically ordered systems, which
are invariant under the combination of lattice translations and spin
rotations. Prominent examples include the chiral spin
density wave states\cite{Martin2008,Li2012a} in $\frac{1}{2}$-doped
triangular lattice systems and $\frac{1}{4}$-doped honeycomb lattice
systems, where $\bar\rho=1/2 \mbox{ (mod 1)}$ per unit cell. Their
magnetic orders preserve magnetic translation
\begin{eqnarray}
\tilde T_{1,2}=(\imth\hat n_{1,2}\cdot\vec\sigma)T_{1,2},~\hat
n_1\cdot\hat n_2=0\Longrightarrow\{\tilde T_1,\tilde T_2\}=0\notag
\end{eqnarray}
Microscopically the spin chirality effectively provides a $\pi$ flux
($\phi=1/2$) per unit cell, and the electronic ground state supports
Chern bands. But due to theorem (\ref{thm}), we know that $\sigma_{xy}$
must be an odd integer independent of microscopic details.

\subsection{Fractional Quantum Hall States on lattice}

Next we discuss the application of (\ref{thm:fractional}) to lattice FQHEs. Once we specify the topological order in the bulk, anyon statistics $\theta_{F,a}$ can only take certain discrete values and (\ref{thm:fractional}) can provide a strong constraint on the Hall conductivity. As a demonstration we work out two examples explicitly: $\nu=1/2$ Laughlin state and $Z_2$ topological order (toric code). According to Theorem \ref{theorem.FQHE} they all satisfy the following conditions:
\begin{eqnarray}
\bar\rho=\tilde\sigma_{xy}\phi+\frac{\theta_{F,a}}{2\pi}\mod1,~~~\tilde\sigma_{xy}=\frac{\theta_{F,F}}{2\pi}\mod1.
\label{thm: Z2 topo order}
\end{eqnarray}

First let's consider $\nu=1/2$ Laughlin state which hosts quasiparticle $s$ obeying semion statistics. The only nontrivial mutual statistics is $\theta_{s,s}=\pi$ and hence there are two different situations. First scenario corresponds to familiar chiral spin liquids~\cite{Kalmeyer1987} and fractional Chern insulators, where the fluxon is a semion $F\sim s$ and hence Hall conductance is a half integer
\begin{eqnarray}
\tilde\sigma_{xy}=\frac12\mod1.
\end{eqnarray}
Without loss of generality, we consider $\tilde\sigma_{xy}=1/2$ and in this case we have
\begin{eqnarray}
\bar\rho=\frac12\phi+\frac{\theta_{s,a}}{2\pi}\mod1\Longrightarrow\bar\rho=\frac12\phi\mod\frac12
\end{eqnarray}
In a usual lattice model with translational symmetry ($\phi=0$), $\bar\rho=\frac{\theta_{s,a}}{2\pi}\mod1$ and therefore only half filling $\bar\rho=\frac12\mod1$ and integer filling are compatible to a $\nu=\frac12$ Laughlin state.

In the 2nd scenario the fluxon is a boson ($F\sim1$), leading to an integer Hall conductance. Since $\theta_{1,a}=0$, we have relation
\bea
\bar\rho=\tilde\sigma_{xy}\phi\mod1,~~~\tilde\sigma_{xy}\in\mbz.
\eea
It can simply be realized in a system where charge-neutral (bosonic) particle-hole excitations form a $\nu=1/2$ Laughlin state on top of charged background. In this scenario only an integer filling is compatible.

Next we comment on toric-code-type\cite{Kitaev2003} $Z_2$ topological
order.  As we will show below, our result has an interesting consequence
even when there is no magnetic field and the Hall conductivity vanishes.
The $Z_2$ topological phase has 3 types of anyons $\{e,m,\epsilon\}$,
which obey mutual semion statistics
$\theta_{e,m}=\theta_{e,\epsilon}=\theta_{m,\epsilon}=\pi$ between
distinct anyons. In this theory, all anyons have a trivial self braiding
angle $\theta_{a,a}=0$. This implies $\theta_{F,F}=0$ and thus
$\tilde{\sigma}_{xy} \in \mathbb{Z}$ because of
Eq.~\eqref{eq.sigmaxy_FF}.
\begin{equation}
 \bar\rho=\tilde\sigma_{xy}\phi+\frac{\theta_{F,a}}{2\pi}\mod1
\end{equation}
In a usual lattice model with $\phi=0$, this further leads to
$\bar\rho=\frac{\theta_{F,a}}{2\pi}\mod1$. This means that at half
filling, both the background anyon and the fluxon are nontrivial anyons,
and they cannot be of the same type. This is exactly the situation for
$Z_2$ spin liquids in a half-filled Mott insulator, where $F\sim m$ and
$a\sim e$.


\section{Conclusions and Discussions}
\label{sec:conclusion}

In this work, we have studied quantum Hall effect in two-dimensional
periodic potentials or periodic lattices. We formulate and prove a
universal relation among the quantized Hall conductivity, the charge and
flux densities per the physical unit cell, and the possible anyon
statistics: Eq.~\eqref{thm} for IQHE, Eq.~\eqref{thm:fractional} for
FQHE, and Eq.~\eqref{thm:nonsymmorphic:fractional} for a non-symmmorphic lattice
with a glide symmetry.  This serves as a strong constraint on the
possible gapped insulating ground state of the system.  While our work
may be regarded as a combination of the Laughlin's gauge argument and
the LSM theorem, our result gives a stronger constraint than what can be
obtained by applying each of these known results independently.  In
particular, according to the standard generalization of the LSM theorem,
a featureless insulator without fractionalization would be possible even
for a fractional filling (\textit{i.e.} fractional number of particles
per physical unit cell) if the number of particle per \emph{magnetic}
unit cell is integer.  Nevertheless, our result implies that these
insulators cannot be a trivial Mott insulator and must have a
nonvanishing Hall conductivity.

Thus, our results constitute useful guides for the
search of desired integer and fractional QH states in a generic
many-body system with magnetic translational symmetries. They provide
strong constraints on the possible gapped ground states at a given
filling number and a given magnetic flux density. They can be applied to
Harper-Hofstadter models and certain magnetically ordered systems.

Our result follows from large gauge invariance and magnetic translation
symmetry, is applicable to general interacting quantum many-particle
systems in periodic potentials.  We developed several different
approaches, which are complementary to each other: (i) a cut and glue
proof relying on edge states, presented in Section~\ref{sec:cut and
glue}; (ii) a pure bulk proof based on the polarization operator,
presented in Section~\ref{sec:IQHE-quantization}; (iii) arguments based
on the Berry phase of moving a $\Phi_0$ fluxon around a closed loop,
presented in Sections~\ref{sec:fluxon_stitching} and \ref{sec:fluxon
braiding}; (iv) arguments based on the symmetry fractionalization
algebra of a $\Phi_0$ fluxon, presented in Section~\ref{sec:frac_mag_trans}.
Each of these arguments may be also useful in exploring other issues in quantum many-body problem.

In this work we have focused on insulators with a global $U(1)$ charge
conservation in two spatial dimensions. Magnetic translations can also
be defined for a discrete global symmetry. In an outlook for the future,
it will be interesting to generalize the framework to systems with other
global symmetries, in other words symmetry protected and enriched
topological phases~\cite{Senthil2014}. Another direction is to generalize
this work to other spatial dimensions.

{As a very interesting development after the present paper
appearing in arXiv,
Matsugatani \textit{et al.}~\cite{Matsugatani-MBChern2018}
proposed an alternative formulation of
the present result in terms of many-body Chern number, and
derived stronger constraints in the presence of crystal symmetries.}

\acknowledgments

We thank Gil-Young Cho, Eduardo Fradkin, Miklos Lajko, Sid Parameswaran, {Nick Read}, T. Senthil, Kirill Shtengel, and Haruki Watanabe for useful discussions.~{We thank Aspen Center for Physics and Kavli Institute for Theoretical
Physics, UC Santa Barbara, for hospitality,
where parts of this
work were performed under support of NSF grants PHY-1066293 and PHY-1607611,
and NSF grants PHY-1125915 and PHY-1748958, respectively.}
This work is
also supported by NSF under award No. DMR-1653769 (YML) and DMR-1151440 (YR), and by MEXT/JSPS KAKENHI Grants
Nos. JP16K05469, JP17H06462, and JP19H01808 (MO).

\appendix

\section{Charge pumping and generalized Luttinger's theorem for edge states of Abelian QHE}\label{app:chiral boson}

First we prove the relation (\ref{translating cylinders}) between charge densities on the edge of two neighboring cylinders. Assuming the bulk has a unique ground state without topological order, its low-energy long-wavelength physics can be captured by a multi-component Abelian Chern-Simons theory\cite{Wen1995}
\begin{eqnarray}\label{bulk cs theory}
\mathcal{L}_\text{bulk}=\frac{\epsilon^{\mu\nu\rho}}{4\pi}\sum_{I,J}{\bf K}_{I,J}a^I_\mu\partial_\nu a^J_\rho-\frac{\epsilon^{\mu\nu\rho}}{2\pi}A_\mu\sum_I {\bf q}_I\partial_\nu a_\rho^I
\end{eqnarray}
where $a_\mu^I$ are dynamical gauge fields describing bulk excitations. ${\bf K}$ is an integer-valued non-singular symmetric matrix satisfying $|\det{\bf K}|=1$, and ${\bf q}$ is an integer-valued vector called ``charge vector''. Gauge invariance on an open manifold leads to the following effective edge theory for the left edge along $\vec a_2\parallel\hat y$-direction
\begin{eqnarray}
&\notag\mathcal{L}_\text{L-edge}=\sum_{I,J}\frac{{\bf K}_{I,J}}{2\pi}\partial_t\phi_I\partial_y\phi_J\\
&+\sum_I{\bf q}_I\frac{A_0\partial_y\phi_I-A_y\partial_0\phi_I}{2\pi}
\end{eqnarray}
where $\phi_I(y,t)$ are chiral boson fields which describes low-energy dynamics of the left edge, satisfying the current algebra
\begin{eqnarray}\label{current algebra}
[\phi_I(y^\prime),\partial_y\phi_J(y)]=2\pi\imth{\bf K}_{I,J}^{-1}\delta(y-y^\prime)
\end{eqnarray}
The charge density on the edge is given by
\begin{eqnarray}
\rho(y)=\frac1{2\pi}\sum_I{\bf q}_I\partial_y\phi_I(y)
\end{eqnarray}
As shown in FIG. \ref{fig:torus glue} the two neighboring cylinders are related by $(\tilde T_1)^{L_1}$ operation, which includes a large gauge transformation
\begin{eqnarray}
\notag&\calU_2^{\phi l_1L_2}\equiv e^{\imth2\pi\phi L_1\sum_{\bf r}r_2\hat N_{\bf r}}\\
&=\exp\big[\imth2\pi\phi L_1\int y\rho(y)\text{d}y\big]
\end{eqnarray}
in addition to pure lattice translations. As a result we have
\begin{eqnarray}
&\notag\rho_{n+1,L}(y)=\calU_{2,n}^{\phi l_1L_2}\rho_{n,L}(y)\calU_{2,n}^{-\phi l_1L_2}\\
&=\rho_{n,L}(y)-\phi l_1\cdot\sigma_{xy},~~~\sigma_{xy}={\bf q}^T{\bf K}^{-1}{\bf q}.
\end{eqnarray}
where $\calU_{2,n}$ denotes an elementary large gauge transformation acting on left edge of $n$-th cylinder. We've used the current algebra (\ref{current algebra}) in the derivation. Similar derivations can be made for the right edge, where the change of charge density is opposite to the left edge.

This is the well-known chiral anomaly\cite{Fujikawa2014B}: \ie chiral edge modes of QHEs is not invariant under large gauge transformations. The chiral anomaly of edge states is canceled\cite{Wen1995} by the anomaly of bulk Chern-Simons theory (\ref{bulk cs theory}), so that the entire system is still gauge invariant.

Next we prove the generalized Luttinger's theorem (\ref{luttinger}) for the edge states of a 2d Abelian insulator. It's straightforward to verify from current algebra (\ref{current algebra}) that the total momentum operator on the edge is written as
\begin{eqnarray}
P_y=\sum_{I,J}\frac{{\bf K}_{I,J}}{4\pi}\int\text{d}y\partial_y\phi_I\partial_y\phi_J
\end{eqnarray}
satisfying
\begin{eqnarray}
[\phi_I(y),P_y]=\imth\partial_y\phi_I(y)
\end{eqnarray}
Now let's insert a flux quantum $\Phi_0$ through the cylinder. It's implemented by large gauge transformation $\calU_{1/L_2}$ and it's straightforward to see that
\begin{eqnarray}
\calU_2~P_y~\calU_2^{-1}=P_y+{2\pi}\int_0^{L_2}\frac{\text{d}y}{L_2}\rho(y)
\end{eqnarray}
Therefore we've proved the generalized Luttinger's theorem (\ref{luttinger}) for the edge states of a gapped bulk.

\section{Berry phase of fluxon braiding in Abelian QHE}\label{app:fluxon braiding}

In this section we compute the Berry phase obtained by adiabatically dragging a $\Phi_0$ fluxon $F$ around a closed contour $\mathcal{C}$ enclosing certain anyon excitations. For any Abelian topological order, the effective Lagrangian describing this adiabatic process is\cite{Wen1995}
\begin{eqnarray}\label{braiding:bare action}
\notag&\mathcal{L}_0=\frac{\epsilon^{\mu\nu\rho}}{4\pi}\sum_{I,J}{\bf K}_{I,J}a^I_\mu\partial_\nu a^J_\rho-\sum_I a_\mu^I(\frac{\epsilon^{\mu\nu\rho}}{2\pi}{\bf q}_I\partial_\nu A_\rho+j^\mu_I)\\
&
\end{eqnarray}
where
\bea
A_\mu({\bf r},t)=\delta A_\mu({\bf r},t)+\bar A_\mu({\bf r})
\eea
is the external (classical) EM vector potential. The time-independent piece $\bar A_\mu({\bf r})$ describes a uniform background magnetic field of strength $\phi\Phi_0$ per unit cell (we've chosen the area of unit cell to be unity for simplicity) as shown in (\ref{app:bg flux}), while the time-dependent piece $\delta A_\mu({\bf r},t)$ is introduced to adiabatically move a fluxon as shown in (\ref{app:bg fluxon}).

Meanwhile $a_\mu^I$ are dynamical gauge fields describing bulk excitations. ${\bf K}$ is an integer-valued non-singular symmetric matrix and ${\bf q}$ is an integer-valued vector called ``charge vector''. The conserved $U(1)$ charge current in 2+1-D is given by
\begin{eqnarray}
J^\mu_{e}=\frac{\epsilon^{\mu\nu\rho}}{2\pi}\sum_I{\bf q}_I\partial_\nu a^I_\rho.
\end{eqnarray}
Meanwhile
\begin{eqnarray}
j_I^\mu({\bf r})=\delta_{\mu,0}j_I^0({\bf r})=\delta_{\mu,0}\sum_{\alpha_I=1}^{N_I}\delta({\bf r}-{\bf r}_{\alpha_I}).
\end{eqnarray}
are time-independent classical fields describing the background anyons located at $\{{\bf r}_{\alpha_I}|1\leq\alpha_I\leq N_I\}$ in the ground state. The fractional statistics of bulk anyon excitations are encoded by the ${\bf K}$ matrix in (\ref{braiding:bare action}). In the adiabatic process, flux insertion and movement won't change the original background anyon $j_I^\mu({\bf r})$. Adiabatic dragging a fluxon along a closed contour $\mathcal{C}$ is implemented by a time-dependent classical vector potential $A_\mu({\bf r},t)$ satisfying
\begin{eqnarray}\label{app:bg fluxon}
&\epsilon^{0\mu\nu}\partial_\mu\delta A_\nu({\bf r},t)=2\pi\delta^{2}\big({\bf r}-{\bf r}_0(t)\big),\\
&\epsilon^{0\mu\nu}\partial_\mu\bar A_\nu({\bf r})=\phi.\label{app:bg flux}
\end{eqnarray}
where the fluxon located at ${\bf r}_0(t)\in\mathcal{C}$ goes along closed contour $\mathcal{C}$ once in time period $0\leq t\leq T$. The Berry phase acquired in this process can be obtained as
\begin{eqnarray}
\int_0^T\text{d}t\int\text{d}^2{\bf r}A_\mu({\bf r},t)J^\mu_e=2\pi\int_{{\bf r}\in\mathcal{C}}\text{d}^2{\bf r}J^0_e=2\pi\rho_{\in\mathcal{C}}
\end{eqnarray}
which is nothing but $2\pi$ times the total charge number in the area enclosed by $\mathcal{C}$. This corresponds to the l.h.s. of Eq. (\ref{eq.tiltilT1_frakF1_phase}), $e^{2\pi\imth\bar\rho L_2}$ when the contour encloses $L_2$ unit cells in one column of the torus.

We can also obtain this Berry phase from the effective response theory for external EM vector potential $\delta A_\mu$, by integrating out all bulk excitations captured by Abelian gauge fields $\{a_\mu^I\}$. This will lead to a Chern-Simons term of the following form
\begin{eqnarray}
&\notag\prod_I\int\mathcal{D}a^I_\mu~e^{-\int\text{d}t\text{d}^2{\bf r}\mathcal{L}_0}=e^{-\int\text{d}t\text{d}^2{\bf r}\mathcal{L}_\text{eff}}\Longrightarrow\\
&\notag\mathcal{L}_{eff}=-\frac{\epsilon^{\mu\nu\rho}}{4\pi}\sum_{I,J}{\bf K}^{-1}_{I,J}\cdot\\
&\notag({\bf q}_IA_\mu-2\pi\frac{\epsilon_{\mu\alpha\beta}}{\partial^2}\partial^\alpha j^\beta_I)\partial_\nu({\bf q}_JA_\rho-2\pi\frac{\epsilon_{\rho\gamma\delta}}{\partial^2}\partial^\gamma j_J^\delta)\\
&=-\frac{\sigma_{xy}}{4\pi}\epsilon^{\mu\nu\rho}\delta A_\mu\partial_\nu\bar A_\rho-\sum_I{\bf q}_I{\bf K}^{-1}_{I,J}\delta A_\mu j_J^\mu,\label{braiding: eff action}\\
&\notag\sigma_{xy}={\bf q}^T{\bf K}^{-1}{\bf q}.
\end{eqnarray}
where we've used relation
\begin{eqnarray}
\partial_\mu j^\mu=0\Longrightarrow\epsilon^{\mu\nu\rho}{\partial_\nu}d_\rho=j^\mu,
~~d_\mu=-\epsilon_{\mu\nu\rho}\frac{\partial^\nu}{\partial^2}j^\rho.
\end{eqnarray}
Note that in (\ref{braiding: eff action}) the Hopf term\cite{Wilczek1983}
\begin{eqnarray}
&\notag\mathcal{L}_{Hopf}=-\pi\sum_{I,J}\epsilon^{\mu\nu\rho}d^I_\mu~{\bf K}^{-1}_{I,J}~j^\mu_J\\
&\notag=-\pi\sum_{I,J}\epsilon^{\mu\nu\rho}d^I_\mu~{\bf K}^{-1}_{I,J}~{\partial_\nu}d^J_\rho\\
&=-\pi\sum_{I,J}\epsilon_{\mu\nu\rho}j_I^\mu~{\bf K}^{-1}_{I,J}~\frac{\partial^\nu}{\partial^2}j_J^\rho\label{braiding:hopf}
\end{eqnarray}
vanishes for stationary background anyons $j_I^\mu({\bf r})=\delta_{\mu,0}j_I^0({\bf r})$. The 1st term in (\ref{braiding: eff action})
\begin{eqnarray}\label{braiding:AB}
\mathcal{L}_{AB}=-\frac{\sigma_{xy}}{4\pi}\epsilon^{\mu\nu\rho}\bar A_\mu\partial_\nu\delta A_\rho
\end{eqnarray}
can be understood as Aharonov-Bohm phase (\ref{AB phase}) between polarization charge $\frac{\epsilon^{\mu\nu\rho}}{2\pi}\partial_\nu A_\rho$ near flux core ${\bf r}_0(t)$ and background flux\cite{GOLDHABER1989}. Compared to Hopf term (\ref{braiding:hopf}), the 2nd term in (\ref{braiding: eff action})
\begin{eqnarray}
&\notag\mathcal{L}_{F,a}=-\sum_{I,J}{\bf q}_I~{\bf K}^{-1}_{I,J}~\delta A_\mu j_J^\mu\\
&=-2\pi\sum_{I,J}\epsilon^{\mu\nu\rho}d^I_\mu~{\bf K}^{-1}_{I,J}~\frac{\partial_\nu \delta A_\rho}{2\pi}{\bf q}_J\label{braiding:fluxon+anyon}
\end{eqnarray}
is nothing but mutual braiding statistics (\ref{phase:anyon}) between fluxon (labeled by quasiparticle current ${\bf q}_I\frac{\epsilon^{\mu\nu\rho}}{2\pi}\partial_\nu A_\rho$) and background anyons (quasiparticle current $j^\mu_I$).

If the bulk has a unique gapped ground state (no topological order), we have $|\det{\bf K}|=1$ and hence the 2nd term is always a multiples of $2\pi$. Thus for Abelian topological orders, we've established the equality between total phase factor on the l.h.s. of (\ref{eq.tiltilT1_frakF1_phase}), and the sum of AB phase (\ref{AB phase}) and anyon braiding phase (\ref{phase:anyon}) on the r.h.s. of (\ref{eq.tiltilT1_frakF1_phase}).

Notice that in the effective field theory (\ref{braiding: eff action}) describing the long-wavelength response of the system to fluxons:
\bea\label{eff lag for braiding}
\mathcal{L}_{eff}=-\delta A_\mu\sum_I{\bf q}_I{\bf K}^{-1}_{I,J}\big[\frac{\epsilon^{\mu\nu\rho}}{4\pi}{\bf q}_J\bar A_\rho+j_J^\mu\big]
\eea
there is an apparent ambiguity on the definition of (time-independent) background vector potential $\bar A_\mu$ and background anyon density $j^\mu_J$, since we can always redefine them in the following way
\bea
\bar A_\mu\rightarrow\bar A_\mu+\alpha_\mu,~~~j_J^\mu\rightarrow j_J^\mu-{\bf q}_J\alpha_\mu.
\eea
without changing the effective Lagrangian (\ref{eff lag for braiding}) at all. This is exactly the ambiguity discussed in (\ref{redef bg flux+anyon}), where the background flux density increase by one flux quantum per unit cell, while the background anyon density changes by removing one fluxon $-F$ per unit cell.


\section{Magnetically translating a unit fluxon}\label{app:mag_trans_fluxon}
Our goal in this section is to establish:
\begin{align}
 \Omega_{\tilde T_1}(F)\Omega_{\tilde T_2}(F)\Omega^{-1}_{\tilde T_1}(F)\Omega^{-1}_{\tilde T_2}(F)=e^{2\pi i\bar\rho},\label{eq:fluxon_mag_trans}
\end{align}
where $\Omega_{\tilde T_i}$ ($i=1,2$) are the fractionalized magnetic translation operators acting on the region $D$, and $\Omega_{\tilde T_i}(F)$ describes the action of this operator on the unit-fluxon-$F$ located inside $D$. Here by definition, \emph{the unit-fluxon-$F$ is a long-length-scale, low-energy object which is created by adiabatically inserting $2\pi$ magnetic flux over a large area inside $D$}.

Let us denote the unknown phase factor $\Omega_{\tilde T_1}(F)\Omega_{\tilde T_2}(F)\Omega^{-1}_{\tilde T_1}(F)\Omega^{-1}_{\tilde T_2}(F)$ as $\eta$. To reach our goal, we firstly show that the phase $\eta$ from this discrete algebra can be computed by an adiabatic Berry's phase corresponding to transporting the unit-fluxon-$F$ around a unit cell. Second, this adiabatic Berry's phase will be computed to be $e^{2\pi i\bar\rho}$.

By locality and the defining property Eq.(\ref{eq:fractionalization}), the conjugation transformation of any local quantum operator $\hat O(D)$ acting within region $D$ by fractionalized operators $\Omega_{\tilde T_i}$ is the same as the conjugation transformation by the global operator $\tilde T_i$. Namely:
\begin{align}
 \Omega_{\tilde T_i}\hat O(D)\Omega_{\tilde T_i}^{-1}=\tilde T_i\hat O(D)\tilde T_i^{-1},\; i=1,2.\label{eq:frac_operator_conj}
\end{align}
From now on we will adopt the following notation for conjugation transformation:
\begin{align}
 {}^{\tilde T_i}\hat O(D)\equiv \tilde T_i\hat O(D)\tilde T_i^{-1}.
\end{align}

Let us denote a quantum state with one fluxon-$F$ inside D as $|F\rangle$. To be precise, $|F\rangle$ also hosts an anti-fluxon somewhere far awary from $D$. We define $\Omega_{\tilde T_i}|F\rangle\equiv|\Omega_{\tilde T_i} F\rangle$. Consider any two local quantum operators $\hat O_{1}$, and $\hat O_{2}$ such that the following quantum amplitudes are nonvanishing:
\begin{align}
 \langle \Omega_{\tilde T_1} F|\hat O_1|F\rangle\neq0;\;\;\langle \Omega_{\tilde T_2} F|\hat O_2|F\rangle\neq0.\label{eq:O_condition}
\end{align}
From Eq.(\ref{eq:frac_operator_conj}), we then know that the relations between quantum amplitudes:
\begin{align}
 \langle \Omega_{\tilde T_1} F|\hat O_1|F\rangle&=\langle \Omega_{\tilde T_2}\Omega_{\tilde T_1} F|{}^{\tilde T_2}\hat O_1|\Omega_{\tilde T_2} F\rangle,\notag\\
 \langle \Omega_{\tilde T_2} F|\hat O_2|F\rangle&=\langle \Omega_{\tilde T_1}\Omega_{\tilde T_2} F|{}^{\tilde T_1}\hat O_2|\Omega_{\tilde T_1} F\rangle.\label{eq:sym_requirement}
\end{align}
Because by definition, $|\Omega_{\tilde T_1}\Omega_{\tilde T_2} F\rangle$ and $|\Omega_{\tilde T_2}\Omega_{\tilde T_1} F\rangle$ can at most differ by a phase factor, which is just $\eta$:
\begin{align}
  |\Omega_{\tilde T_1}\Omega_{\tilde T_2} F\rangle=\eta| \Omega_{\tilde T_2}\Omega_{\tilde T_1} F\rangle.
\end{align}
Namely $\eta$ can be computed as the phase of the following product:
\begin{align}
 \eta\sim & \langle F|\hat O_2^{-1}|\Omega_{\tilde T_2} F\rangle\cdot\langle\Omega_{\tilde T_2} F |{}^{\tilde T_2}(\hat O_1^{-1}) |\Omega_{\tilde T_1}\Omega_{\tilde T_2} F\rangle \notag\\
 &\cdot\langle \Omega_{\tilde T_1}\Omega_{\tilde T_2} F|{}^{\tilde T_1}\hat O_2|\Omega_{\tilde T_1} F\rangle\cdot \langle \Omega_{\tilde T_1} F|\hat O_1|F\rangle \label{eq:eta_prod}
\end{align}
The advantage of this expression of $\eta$ is that it is explicitly independent of the global phase choices of involved four fluxon states. Although Eq.(\ref{eq:eta_prod}) holds for arbitrary operators $\hat O_i$ satisfying Eq.(\ref{eq:O_condition}), there exist particularly convenient choices of these operators. Let us choose $\hat O_i=\hat O^{ad}_i$ as the adiabatic transporting of the unit fluxon $|F\rangle$ from its original position by one lattice spacing along the $\vec a_i$ direction ($i=1,2$). Precisely, $\hat O_i^{ad}$ are defined as the following time-revolutions:
\begin{align}
 \hat O_i^{ad}=W_i\cdot \mathcal{T}\; exp\big(-i\int_0^T dt H_{i}(t)\big),
\end{align}
where $\mathcal{T}$ is the time ordering. Starting from the original Hamiltonian $H_{orig}$,  $H_{i}(0)\equiv H_0$ is the modified Hamiltonian whose ground state hosts a fluxon $|F\rangle$ located in the region $D$ (and another anti-fluxon far away from $D$). $W_i H_{i}(T) W_i^{-1}={}^{\Omega_{\tilde T_i}}H_0$ which hosts the fluxon-$F$ in $D$ spatially translated by one lattice spacing (but the anti-fluxon is not movied), and $W_i$ is the local gauge transformation near $F$ needed in order to make $H_{i}(t)$ a smooth time-evolution.

Here the definition of ${}^{\Omega_{\tilde T_i}}H_0$ is the following: for those terms of $H_0$ near the location of the fluxon $F$, they are conjugated by the global operation $\tilde T_i$; while all other terms stay invariant as in $H_0$. Because inside region $D$, $H_0$ is different from $H_{orig}$ only by those terms near the location of $F$, and $H_{orig}$ is $\tilde T_i$ symmetric, such a definition is self-consistent. Clearly, $|\Omega_{\tilde T_i}F\rangle$ is the ground state of ${}^{\Omega_{\tilde T_i}}H_0$. Note that ${}^{\Omega_{\tilde T_1}\Omega_{\tilde T_2}}H_0={}^{\Omega_{\tilde T_2}\Omega_{\tilde T_1}}H_0$ (due to the fact that every local term of Hamiltonian need to be $U(1)$ symmetric) and |$\Omega_{\tilde T_1}\Omega_{\tilde T_2} F\rangle$ is its ground state.

We will adopt the notation ${}^{\Omega_{\tilde T_i}}\hat S$ for a similarly defined transformation of a general operator $\hat S$ which is a summation of many local terms, which requires that within $D$, $^{\tilde T_i}\hat S-\hat S$ is composed of terms only near the location of the fluxon $F$. After discretizing the time-evolution then considering the limit of the time-step going to zero, it is straightforward to establish that:
\begin{align}
\langle \Omega_{\tilde T_1} F|\hat O^{ad}_1|F\rangle&=\langle \Omega_{\tilde T_2}\Omega_{\tilde T_1} F|{}^{\Omega_{\tilde T_2}}\hat O^{ad}_1|\tilde T_2 F\rangle,\notag\\
 \langle \Omega_{\tilde T_2} F|\hat O^{ad}_2|F\rangle&=\langle \Omega_{\tilde T_1}\Omega_{\tilde T_2} F|{}^{\Omega_{\tilde T_1}}\hat O^{ad}_2|\tilde T_1 F\rangle,
\end{align}
where ${}^{\Omega_{\tilde T_j}}\hat O^{ad}_i={}^{T_j}W_i\cdot \mathcal{T}\; exp\big(-i\int_0^T dt {}^{\Omega_{\tilde T_j}}H_{i}(t)\big)$. Note that because gauge transformations commute with each other, we have: ${}^{\tilde T_j}W_i={}^{T_j}W_i$; namely the conjugate of magnetic translation on gauge transformation is just as the usual translation. This identity is eventually related to the fact that the magnetic translation algebra of the unit fluxon, just as the usual translation algebra, measures $e^{2\pi i\bar\rho}$.

Therefore, $\eta$ can be computed as:
\begin{align}
 \eta\sim & \langle F| \mathcal{T}\; exp\big(i\int_{-T}^0 dt H_{2}(-t)\big)\cdot W_2^{-1}\notag\\
 &\cdot \mathcal{T}\; exp\big(i\int_{-T}^0 dt {}^{\Omega_{\tilde T_2}}H_{1}(-t)\big)\cdot{}^{T_2}W_1^{-1}\notag\\
 &\cdot {}^{T_1}W_2\cdot \mathcal{T}\; exp\big(-i\int_0^T dt {}^{\Omega_{\tilde T_1}}H_{2}(t)\big) \notag\\
 &\cdot W_1\cdot \mathcal{T}\; exp\big(-i\int_0^T dt H_{1}(t)\big)|F\rangle\notag\\
 &=\langle F| {W_2^{-1}{}^{T_2}W_1^{-1}{}^{T_1}W_2W_1}\notag\\
 &\cdot {}^{W_1^{-1}{}^{T_1}W_2^{-1}{}^{T_2}W_1W_2}[\mathcal{T}\; exp\big(i\int_{-T}^0 dt H_{2}(-t)\big)]\notag\\
 &\cdot {}^{W_1^{-1}{}^{T_1}W_2^{-1}{}^{T_2}W_1}[\mathcal{T}\; exp\big(i\int_{-T}^0 dt {}^{\Omega_{\tilde T_2}}H_{1}(-t)\big)]\notag\\
 &\cdot  {}^{W_1^{-1}}[\mathcal{T}\; exp\big(-i\int_0^T dt {}^{\Omega_{\tilde T_1}}H_{2}(t)\big)] \notag\\
 &\cdot \mathcal{T}\; exp\big(-i\int_0^T dt H_{1}(t)\big)|F\rangle\notag\\
 &\equiv\langle F|{W_2^{-1}{}^{T_2}W_1^{-1}{}^{T_1}W_2W_1}\cdot \hat O^{ad}_d\hat O^{ad}_l\hat O^{ad}_u\hat O^{ad}_r|F\rangle\label{eq:cumulative}
\end{align}
After the second equation mark, we use accumulated gauge transformations to conjugate the time-evolutions. After this treatment, the gauge during the whole four-step time-evolution is smooth, and we are only left with the final gauge transformation $W_2^{-1}{}^{T_2}W_1^{-1}{}^{T_1}W_2W_1$, which is easy to compute. After the third equation mark, we reserve the symbols $\hat O^{ad}_d,\hat O^{ad}_l,\hat O^{ad}_u,\hat O^{ad}_r$ to describe the adiabatic time-evolutions for the right,up,left and down move in this smooth gauge.

Next we compute this phase $\eta$ explicity by perturbation theory. The basic idea is to choose the gauge in $H_{1}(t),H_2(t)$ so that they are only perturbatively different from $H_0$. We choose a rectangular region on a square lattice to demonstrate the calculation. Let $(x,y)$ be the coordinate of a site. The adjacent sites are connected by links carrying the gauge field: $A_x(x,y)$ connect $(x,y)$ to $(x+1,y)$, and $A_y(x,y)$ connect $(x,y)$ to $(x,y+1)$. Consider a rectangular region: $0\leq x\leq L_x,0\leq y\leq L_y$ which contains $L_xLy$ plaquettes. Initially, the state $|F\rangle$ is the ground state of $H_0$ hosting a single flux uniformly distributed in this region (i.e. every plaquette hosts $\varphi\equiv \frac{2\pi}{L_xL_y}$ flux), the corresponding gauge field configuration are denoted as $A^0_{x,y}(x,y)$. One can choose a gauge to describe this flux distribution. Then, we specifically define $H_i(t)$ ($0\leq t\leq 1$) as:
\begin{align}
 H_1(t)=H_0(\{A^{0}_x(x,y),A^0_y(x,y)+\delta A_y(x,y,t)\}),\notag\\
 H_2(t)=H_0(\{A^{0}_x(x,y)+\delta A_x(x,y,t),A^0_y(x,y)\}),
\end{align}
where $\delta A_y(x,y,t)=-\varphi(t)$ if $1\leq x\leq L_x, 0\leq y\leq L_y-1$, and zero otherwise, $\varphi(t=0)=0$ and $\varphi(t=T)=\varphi$. And $\delta A_y(x,y,t)=\varphi(t)$ if $0\leq x\leq L_x-1, 1\leq y\leq L_y$. Namely, we choose a convenient gauge to describe the time-evolutions. When the region is very large, $\varphi$ goes to zero. \emph{We will use $\varphi$ as the perturbative parameter to control the calculation}.

We will only keep the leading order in our perturbative calculation. Taylor expanding these time-dependent Hamiltonians lead to:
\begin{align}
 H_1(t)&=H_0-\varphi(t)\sum_{1\leq x\leq L_x}^{0\leq y\leq L_y-1}j_y(x,y)+O(\varphi^2),\notag\\
 H_2(t)&=H_0+\varphi(t) \sum_{0\leq x\leq L_x-1}^{1\leq y\leq L_y}j_x(x,y)+O(\varphi^2),
\end{align}
where the current operators living on links are obtained by partial differentiating $H_0$ w.r.t. the corresponding gauge fields. It then follows that to the leading order of $\varphi$:
\begin{align}
 \hat O^{ad}_{r/u}&=\mathcal{T}\mbox{exp}\big[- i \int _0^T dt (H_0+O(\varphi) \sum_{0\leq x\leq L_x}^{0\leq y\leq L_y}j_{x}\notag\\
 &+O(\varphi) \sum_{0\leq x\leq L_x}^{0\leq y\leq L_y}j_{y})\big],\notag\\
 \hat O^{ad}_{l/d}&=\mathcal{T}\mbox{exp}\big[+ i \int _0^T dt (H_0+O(\varphi) \sum_{0\leq x\leq L_x}^{0\leq y\leq L_y}j_{x}\notag\\
 &+O(\varphi) \sum_{0\leq x\leq L_x}^{0\leq y\leq L_y}j_{y})\big].
\end{align}
Because we are considering a charge insulator, the spatial summation current density is zero in the ground state of $H_{orig}$. Due to the extra factor $O(\varphi)$, it is safe to ignore these current contributions to the leading order of $\varphi$. We conclude that to the leading order of $\varphi$, $\eta$ is due to the following phase:
\begin{align}
 \eta=\langle F|{W_2^{-1}{}^{T_2}W_1^{-1}{}^{T_1}W_2W_1}|F\rangle,
\end{align}
where ${W_2^{-1}{}^{T_2}W_1^{-1}{}^{T_1}W_2W_1}$ is the required gauge transformation to send the final Hamiltonian $H_{final}$ in the smooth gauge (i.e., the Hamiltonian after $O^{ad}_d$) back to the original gauge of $H_0$. Note that after $O^{ad}_r$, $H=H_0-\varphi\sum_{1\leq x\leq L_x}^{0\leq y\leq L_y-1}j_y(x,y)$. After $O^{ad}_u$, $H=H_0-\varphi\sum_{1\leq x\leq L_x}^{0\leq y\leq L_y-1}j_y(x,y)+\varphi \sum_{1\leq x\leq L_x}^{1\leq y\leq L_y}j_x(x,y)$. After $O^{ad}_l$, $H=H_0-\varphi\sum_{1\leq x\leq L_x}^{0\leq y\leq L_y-1}j_y(x,y)+\varphi \sum_{1\leq x\leq L_x}^{1\leq y\leq L_y}j_x(x,y)+\varphi\sum_{1\leq x\leq L_x}^{1\leq y\leq L_y}j_y(x,y)$. Therefore, after $O^{ad}_d$,
\begin{align}
H_{final}&=H_0-\varphi\sum_{1\leq x\leq L_x}^{0\leq y\leq L_y-1}j_y(x,y)+\varphi \sum_{1\leq x\leq L_x}^{1\leq y\leq L_y}j_x(x,y)\notag\\
 &+\varphi\sum_{1\leq x\leq L_x}^{1\leq y\leq L_y}j_y(x,y)-\varphi \sum_{0\leq x\leq L_x-1}^{1\leq y\leq L_y}j_x(x,y)\notag\\
 &=H_0-\varphi\sum_{1\leq x\leq L_x}^{y=0}j_y(x,y)+\varphi \sum_{x=L_x}^{1\leq y\leq L_y}j_x(x,y)\notag\\
 &+\varphi\sum_{1\leq x\leq L_x}^{y=L_y}j_y(x,y)-\varphi \sum_{x=0}^{1\leq y\leq L_y}j_x(x,y).
\end{align}
Note that the current terms form a loop enclosing  sites $1\leq x\leq L_x, 1\leq y\leq L_y$. Clearly, the required gauge transformation is the following:
\begin{align}
 \eta&= \langle F|W_2^{-1}{}^{T_2}W_1^{-1}{}^{T_1}W_2W_1|F\rangle\notag\\
 &=\langle F|\exp{\big[i \varphi \sum_{1\leq x\leq L_x}^{1\leq y\leq L_y} \hat n(x,y)\big]}|F\rangle,
\end{align}
where $\hat n(x,y)$ is the particle number operator on site $(x,y)$. Because $\varphi=2\pi/(L_xL_y)$, this is exactly measuring the average charge density per unit cell: $e^{2\pi i \bar \rho}$.


%

\end{document}